\documentclass[preprint2,twocolappendix]{aastex6}

\usepackage{natbib}

\usepackage[ampersand]{easylist}

\newcommand{\teff}{T_{\rm eff}}
\newcommand{\feh}{\rm[Fe/H]}

\newcommand{\dnu}{\Delta\nu}
\newcommand{\num}{\nu_\mathrm{max}}
\newcommand{\runo}{r_{010}}
\newcommand{\rdos}{r_{02}}

\shorttitle{The {\it Kepler} Dwarfs asteroseismic LEGACY sample II}
\shortauthors{V.~Silva Aguirre et al.}

\begin{document}

\title{Standing on the shoulders of Dwarfs: the {\it Kepler} asteroseismic LEGACY sample II --- radii, masses, and ages}
\author{V\'ictor Silva Aguirre\altaffilmark{1}}
\author{Mikkel~N.~Lund\altaffilmark{2,1}}
\author{H.~M.~Antia\altaffilmark{3}}
\author{Warrick~H.~Ball\altaffilmark{4,5}}
\author{Sarbani~Basu\altaffilmark{6}}
\author{J\o rgen~Christensen-Dalsgaard\altaffilmark{1}}
\author{Yveline~Lebreton\altaffilmark{7,8}}
\author{Daniel~R.~Reese\altaffilmark{9,2,1}}
\author{Kuldeep~Verma\altaffilmark{3,1}}
\author{Luca~Casagrande\altaffilmark{10}}
\author{Anders~B.~Justesen\altaffilmark{1}}
\author{Jakob~R.~Mosumgaard\altaffilmark{1}}
\author{William~J.~Chaplin\altaffilmark{2,1}}
\author{Timothy~R.~Bedding\altaffilmark{11,1}}
\author{Guy~R.~Davies\altaffilmark{2,1}}
\author{Rasmus~Handberg\altaffilmark{1}}
\author{G\"unter~Houdek\altaffilmark{1}}
\author{Daniel~Huber\altaffilmark{11,12,1}}
\author{Hans~Kjeldsen\altaffilmark{1}}
\author{David~W.~Latham\altaffilmark{13}}
\author{Timothy~R.~White\altaffilmark{1}}
\author{Hugo~R.~Coelho\altaffilmark{2}}
\author{Andrea~Miglio\altaffilmark{2}}
\and
\author{Ben~Rendle\altaffilmark{2}}


\altaffiltext{1}{Stellar Astrophysics Centre, Department of Physics and Astronomy, Aarhus University, Ny Munkegade 120, DK-8000 Aarhus C, Denmark}
\altaffiltext{2}{School of Physics \& Astronomy, University of Birmingham, Edgbaston Park Road, West Midlands, Birmingham, UK, B15 2TT}
\altaffiltext{3}{Tata Institute of Fundamental Research, Homi Bhabha Road, Mumbai 400005, India}
\altaffiltext{4}{Institut f\"ur Astrophysik, Georg-August-Universit\"at G\"ottingen, Friedrich-Hund-Platz 1, 37077, G\"ottingen, Germany}
\altaffiltext{5}{Max-Planck-Institut f\"ur Sonnensystemforschung, Justus-von-Liebig-Weg 3, 37077 G\"ottingen, Germany}
\altaffiltext{6}{Department of Astronomy, Yale University, PO Box 208101, New Haven, CT 06520-8101, USA}
\altaffiltext{7}{Observatoire de Paris, GEPI, CNRS UMR 8111, F-92195 Meudon, France}
\altaffiltext{8}{Institut de Physique de Rennes, Universit\'e de Rennes 1, CNRS UMR 6251, F-35042 Rennes, France}
\altaffiltext{9}{LESIA, Observatoire de Paris, PSL Research University, CNRS, Sorbonne Universit\'es, UPMC Univ. Paris 06, Univ. Paris Diderot,
Sorbonne Paris Cit\'e, 92195 Meudon, France}
\altaffiltext{10}{Research School of Astronomy and Astrophysics, Mount Stromlo Observatory, The Australian National University, ACT 2611, Australia}
\altaffiltext{11}{Sydney Institute for Astronomy (SIfA), School of Physics, University of Sydney, NSW 2006, Australia}
\altaffiltext{12}{SETI Institute, 189 Bernardo Avenue, Mountain View, CA 94043, USA}
\altaffiltext{13}{Harvard-Smithsonian Center for Astrophysics, 60 Garden Street Cambridge, MA 02138 USA}

\begin{abstract}
We use asteroseismic data from the {\it Kepler} satellite to determine fundamental stellar properties of the 66 main-sequence targets observed for at least one full year by the mission. We distributed tens of individual oscillation frequencies extracted from the time series of each star among seven modelling teams who applied different methods to determine radii, masses, and ages for all stars in the sample. Comparisons among the different results reveal a good level of agreement in all stellar properties, which is remarkable considering the variety of codes, input physics and analysis methods employed by the different teams. Average uncertainties are of the order of $\sim$2\% in radius, $\sim$4\% in mass, and $\sim$10\% in age, making this the best-characterised sample of main-sequence stars available to date. Our predicted initial abundances and mixing-length parameters are checked against inferences from chemical enrichment laws  $\Delta Y / \Delta Z$ and predictions from 3D atmospheric simulations. We test the accuracy of the determined stellar properties by comparing them to the Sun, angular diameter measurements, Gaia parallaxes, and binary evolution, finding excellent agreement in all cases and further confirming the robustness of asteroseismically-determined physical parameters of stars when individual frequencies of oscillation are available. Baptised as the {\it Kepler} dwarfs LEGACY sample, these stars are the solar-like oscillators with the best asteroseismic properties available for at least another decade. All data used in this analysis and the resulting stellar parameters are made publicly available for the community.
\end{abstract}
\section{Introduction}\label{s:intro}
%
Asteroseismology has changed the way we determine properties of stars. The detection of oscillations at the surface is a window to their interiors as they allow us to determine their physical properties with a precision that is otherwise extremely hard to achieve in field stars. These properties, ages in particular, are crucial to our understanding of different fields of astrophysics \citep[such as Galactic Archaeology, see e.g.,][]{Miglio:2013hh,2016MNRAS.455..987C}. Stellar age determinations, by traditional means, can be very uncertain \citep[see e.g.,][]{Soderblom:2010kr} and most age estimates, even when they claim to be precise, are not necessarily accurate.

The most widely used method to obtain ages of stars is to compare their observed atmospheric properties with those predicted by theoretical evolutionary sequences \citep[e.g.,][]{Edvardsson:1993uk,1998A&A...329..943N,Pont:2004fw} or theoretical isochrones in the case of star clusters. This approach works reasonably well for clusters as long as the colour-magnitude diagram is well defined and contains stars in all stages of evolution, in particular the main-sequence, the main-sequence turnoff and the red-giant branch. However, the technique of evolutionary-track or isochrone fitting does not work very well for single stars. More sophisticated statistical treatments, particularly Bayesian estimates \citep[e.g.,][]{Jorgensen:2005bn,2007ApJS..168..297T}, are needed to obtain unbiased ages and, even then, uncertainties can be large. Estimating stellar ages also becomes increasingly problematic for stars that are more distant and hence fainter: reddening is uncertain, and the distances to the stars are not yet well known (although Gaia will drastically change this picture).

The CoRoT and {\it Kepler} missions have completely changed the field of asteroseismology. Prior to these missions, solar-like oscillations for stars other than the Sun were observed in only a small number of them \citep[see e.g., ][for a review]{Chaplin:2013gz}. Space-based photometry has now made asteroseimic analyses almost routine. The {\it Kepler} mission detected solar-like oscillations in hundreds of main-sequence and subgiant stars, and the global asteroseismic parameters of these targets have been used to determine properties such as mass and radius \citep{Chaplin:2014jf}. While the mass and radius estimates of these stars are reliable, their ages are less so due to the lack of sensitivity of these global properties to the deep stellar layers, where most of the evolutionary change takes place. To overcome this and determine robust asteroseismic ages for these stars, it is necessary to extract and model the full spectrum of individual oscillation modes.

\citet{2015MNRAS.452.2127S} and \citet{2016MNRAS.456.2183D} analysed and modelled a sample of 33 exoplanet hosts for which the {\it Kepler} mission had obtained asteroseismic data, including individual oscillation frequencies. Those results have informed other studies of exoplanet systems \citep[e.g.,][]{2015ApJ...808..126V,Bonfanti:2016iq,JontofHutter:2016ch}, and of the relation between stellar rotation and age \citep{vanSaders:2016cr}. In this paper we perform a similar analysis to that of \citet{2015MNRAS.452.2127S} for a set of 66 main-sequence stars observed by {\it Kepler} that are not known to be exoplanet hosts. Instead, the target stars are selected as those with the highest signal-to-noise ratios and the most precisely determined oscillation frequencies. In recent years there have been similar studies using smaller samples and observation of shorter duration \citep[e.g.,][]{Appourchaux:2012kd,Mathur:2012bj,Metcalfe:2014ig}. The dataset used for this work has been extracted from the full available set of observations made by the {\it Kepler} mission, which for most stars means frequencies obtained from a four-year time series. Thus, the targets comprising this study, which we baptise as the {\it Kepler} dwarfs LEGACY sample, have the best asteroseismic data available among solar-like stars for at least another decade (i.e., until the PLATO 2.0 mission, \citet{Rauer:2014kx}).

The companion paper by Lund et al.~2016 (submitted) describes the details of the data analysis and the frequencies that have been used in the modelling presented in this study. The rest of this paper is organised as follows. In Section~\ref{s:sample} we describe the selection criteria for choosing the sample of stars that have been analysed. We also discuss the ancillary data such as effective temperatures and metallicity needed for the analysis. The different techniques used in modelling the stars are described in Section~\ref{s:methods}. Section~\ref{s:res} is devoted to the results of the analysis, and tests of the derived properties using independent constraints are given in Section~\ref{s:checks}. We summarise the results and mention some of their applications in Section~\ref{s:conc}.
\section{The LEGACY sample} \label{s:sample}
Our goal was to select main-sequence stars of the highest asteroseismic data quality present in the {\it Kepler} sample. A thorough description of the selection process and the main observational properties of our targets are given in the accompanying paper by Lund et al.~2016 (submitted). We give a brief account of the most important points in this section.

The 66 stars comprising the LEGACY sample were chosen from more than 500 main-sequence and subgiant targets in which {\it Kepler} detected oscillations \citep{Chaplin:2014jf}. We selected all targets that had more than one year of short-cadence observations, and where inspection of the power spectrum did not reveal any clear signature of bumped $\ell=1$ modes. These are modes of mixed character that behave like pressure modes in the envelope and buoyancy modes in the core, and their appearance is related to the end of the core-hydrogen burning phase \citep{Aizenman:1977wh}.
\begin{figure}[ht!]
\centering
\includegraphics[width=\linewidth]{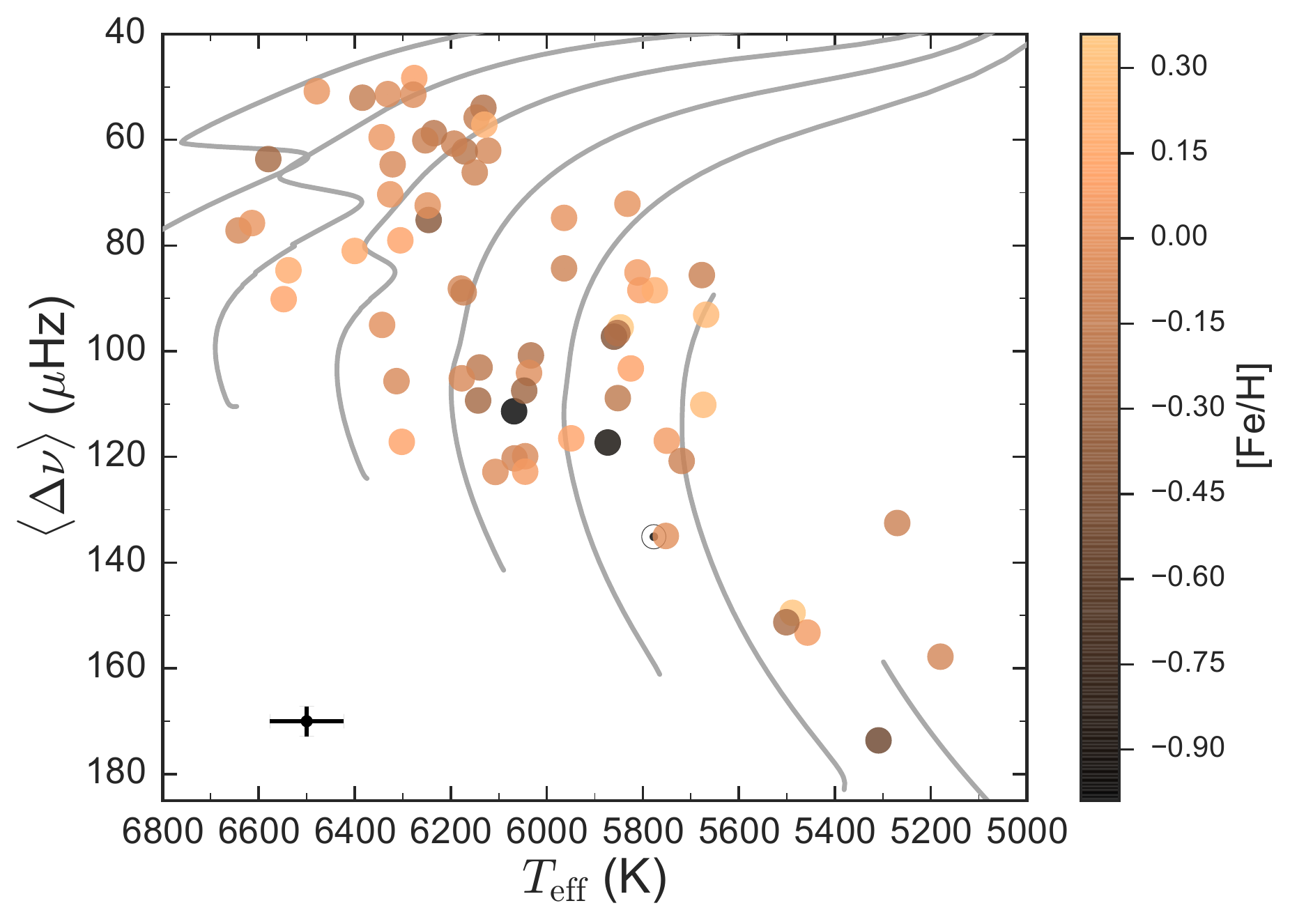}
\caption{Position of the LEGACY sample (filled circles) and the Sun in an asteroseismic HRD, colour coded according to their surface chemical composition. Typical error bar in effective temperature and large frequency separation is shown in black. Also depicted are representative stellar evolution tracks at solar metallicity spanning a mass range between 0.8$-$1.4~M$_\odot$ in steps of 0.1~M$_\odot$ \citep[from][]{2015MNRAS.452.2127S}. As a side note we mention that the two lowest mass tracks do not evolve further into the red giant phase as the computations where stopped at an age of 16 Gyr. \label{fig:seishrd}}
\end{figure}

Figure~\ref{fig:seishrd} shows an asteroseismic Hertzsprung-Russell diagram (HRD) where the luminosity on the y-axis has been replaced by an asteroseismic property known as the average large frequency separation. This quantity decreases as stars evolve, since it is approximately proportional to the square root of the mean stellar density \citep{Ulrich:1986ge,1987Natur.326..257G}:
\begin{equation}
\centering
{\langle\dnu\rangle} \propto \left(\frac{M}{M_\odot}\right)^{1/2}\left(\frac{R}{R_\odot}\right)^{-3/2}\,.
\label{eq:dnuscal}
\end{equation}

The large frequency separation in Fig.~\ref{fig:seishrd} has been determined for the data using a gaussian-weighted linear fit to the $\ell=0$ frequencies as a function of radial order as described in Lund et al.~2016 (submitted). Tracks at solar metallicity are shown in the figure as a reference, depicting the evolution of solar-type stars in this parameter space. Our sample predominantly comprises stars hotter and more evolved than the Sun, an expected bias since the amplitudes of oscillations scale approximately with stellar luminosity \citep{Kjeldsen:1995tr}, making their detection easier in that region of the asteroseismic HRD. The atmospheric composition of our targets spans values from super-solar metallicity at $\feh=0.35$ to metal-poor stars of $\feh=-1.00$, making this sample the most extended one in terms of heavy-element abundances where a homogeneous asteroseismic analysis is feasible using individual frequencies of oscillation. A summary of the global asteroseismic quantities and atmospheric properties of the sample as compiled for this modelling effort is given in Appendix~\ref{app:input}.
\section{Asteroseismic modelling pipelines} \label{s:methods}
Seven independent pipelines were used to determine the stellar properties of the LEGACY sample. These pipelines use a variety of evolutionary and pulsation codes and consider different sets of input physics when computing theoretical evolutionary sequences and oscillation frequencies. Moreover, each pipeline uses its own preferred set of asteroseismic observables and statistical approach to analyse the goodness of fit of the models and extract the stellar properties of each target. In the following we give a brief general description of the common characteristics in the process of estimating stellar properties from asteroseismic modelling, and give the specific details of each pipeline in an itemised list below.

Firstly, one must generate a set of stellar models for different masses and chemical compositions using a stellar evolution code and set of input physics. Next, for many of these models theoretical oscillation frequencies are computed using a standard (adiabatic) pulsation code. The actual number of evolutionary models and frequencies calculated is either defined in advance (when pipelines use precomputed grids of models) or determined during the fitting process (when pipelines chose masses and composition as part of the optimisation). In both cases, the net result is stellar structures with atmospheric properties and theoretically computed frequencies that can be compared to those obtained from observations.

All pipelines use the atmospheric parameters $\teff$ and $\feh$ as constraints but there are variations in the way asteroseismic data are used when determining stellar properties. One method is to fit the individual oscillation frequencies, $\nu_{i}(n)$, using a suitable prescription to correct for the surface effect \citep[see e.g.,][for a physical motivation and description of these corrections]{Kjeldsen:2008kw,Ball:2014gx,Sonoi:2015gy}. Another method is to fit combinations of frequencies, normally the frequency separation ratios, defined as \citep{Roxburgh:2003bb}:
\begin{equation}\label{eqn:r02}
r_{02}(n)=\frac{d_{02}(n)}{\Delta\nu_{1}(n)}
\end{equation}
\begin{eqnarray}\label{eqn:rat}
r_{01}(n)=\frac{d_{01}(n)}{\Delta\nu_{1}(n)},& & r_{10}(n) = \frac{d_{10}(n)}{\Delta\nu_{0}(n+1)}\,.
\end{eqnarray}
Here, $\Delta\nu_{\ell}(n)=\nu_{n,\ell}-\nu_{n-1,\ell}$ is the large separation between modes of same angular degree and consecutive overtone, $d_{02}(n)=\nu_{n,0}-\nu_{n-1,2}$ is the small frequency separation, and $d_{01}(n)$ and $d_{10}(n)$ are the 5-point small frequency separations:
\begin{equation}\label{eqn:d01}
d_{01}(n)=\frac{1}{8}(\nu_{n-1,0}-4\nu_{n-1,1}+6\nu_{n,0}-4\nu_{n,1}+\nu_{n+1,0})
\end{equation}
\begin{equation}\label{eqn:d10}
d_{10}(n)=-\frac{1}{8}(\nu_{n-1,1}-4\nu_{n,0}+6\nu_{n,1}-4\nu_{n+1,0}+\nu_{n+1,1})\,.
\end{equation}

The frequency separation ratios have been shown to be mostly sensitive to the deep layers of the star \citep[see e.g.,][]{Roxburgh:2005hs,OtiFloranes:2005ii,SilvaAguirre:2011jz}, effectively diminishing the impact of poor modelling of the stellar outer layers in 1D evolutionary codes as well as the requirement of line-of-sight velocity corrections \citep[see,][]{Davies:2014fe}. Due to the strong correlations between combinations including five individual frequencies, the ratios are customarily written and reproduced as one unique set of observables, called $\runo$ \citep[see, e.g.,][]{SilvaAguirre:2013in}:
\begin{equation}\label{eqn:r010}
r_{010}=\{r_{01}(n),r_{10}(n),r_{01}(n+1),r_{10}(n+1),...\}\,.
\end{equation}
The pipelines participating in this study use individual frequencies, or combinations of frequencies, or both, following their own preference.

The process of determining stellar properties can be computationally expensive, since some methods produce individual grids for each of the fitted targets that can require calculating thousands of models and frequencies for a given set of input physics. The inclusion of additional physical processes such as microscopic diffusion or overshooting effectively increases the grid-dimensions to explore and can lead to a rapid increase in the number of models required to properly sample the parameter space. One alternative used by some pipelines to reduce the computational load is to use the average large frequency separation $\langle\dnu\rangle$ and the frequency of maximum power in the oscillation spectrum $\num$ as physically motivated proxies to constrain the parameter space that needs to be covered by the models. Since these quantities are sensitive to the mean stellar density and surface gravity, respectively \citep{Ulrich:1986ge,Brown:1991cv}, they can restrict the mass and radius combinations that need to be explored.

All modelling teams were provided with the global asteroseismic parameters ($\langle\dnu\rangle$ and $\num$), the individual frequencies, frequency ratios, and correlations as determined by Lund et al.~2016 (submitted), as well as the atmospheric properties $\teff$ and $\feh$ compiled from the literature. For most of our targets, the effective temperature and composition were taken from the spectroscopic analysis made by \citet{Buchhave:2015cg}, and for the remainder we opted for other sources compiled from the literature. Table~\ref{tab:input} reports the values given to the modelling teams, including the references for the atmospheric properties used.

In the following, we describe the main characteristics of the seven pipelines employed in the analysis and enclose a summary of them in Table~\ref{tab:pipelines}.
\begin{splitdeluxetable*}{cccccBccc}
\tablecaption{Summary of codes, input physics and optimisation methods applied by each pipeline.\label{tab:pipelines}}
\tablehead{
\colhead{}& \colhead{{\ttfamily AIMS}} & \colhead{{\ttfamily ASTFIT}} & \colhead{{\ttfamily BASTA}}  & \colhead{{\ttfamily C2kSMO}} & \colhead{{\ttfamily GOE}} & \colhead{{\ttfamily V\&A}} & \colhead{{\ttfamily YMCM}}
}
\startdata
Models & MESA & ASTEC & GARSTEC & Cesam2k & MESA & MESA & YREC \\
Frequencies & InversionKit & ADIPLS & ADIPLS& LOSC & ADIPLS & ADIPLS & AB94\\
Solar Mixture & GN93 & GN93 & GS98 & GN93 & GS98 &GS98 & GS98 \\
Opacities &OPAL96$+$JF05  & OPAL96$+$JF05 & OPAL96$+$JF05 & OPAL96$+$JF05 & OPAL96$+$JF05 & OP05$+$JF05 & OPAL96$+$JF05 \\
EOS & OPAL05 & OPAL05 & OPAL05 & OPAL05 & OPAL05 & OPAL05 & OPAL05 \\
Nuclear Reactions & NACRE & NACRE & NACRE & NACRE & NACRE & NACRE & Solar Fusion \\
Atmosphere & Eddington grey & Eddington grey & Eddington grey & Eddington grey & Eddington grey & Eddington grey &  Eddington grey \\
Diffusion & No & MP93, $\leq$1.1~M$_\odot$ & T94, $\leq$1.2~M$_\odot$ & MP93 & T94 & T94, $\leq$1.35~M$_\odot$ & No \\
Overshoot & Yes & No & Yes& Yes & Yes & Yes & Yes \\
Convection & CG68 & MTL58 & MLT12 & CGM96 & CG68 & MLT58 & MTL58 \\
$\alpha_\mathrm{conv}$ & 1.8 & 1.5, 1.8, 2.1 & 1.791& Variable & Variable & Variable & Variable \\
$\Delta$Y/$\Delta$Z & 2.0 & 1.0-2.0 & 1.4 & Variable & Variable & Variable & Variable \\
Fitted data & $\nu_i(n)$ & $\nu_i(n)$ & $\runo$, $\rdos(n)$ & $\runo$, $\rdos(n)$, $\nu_0(n_\mathrm{min})$ & $\nu_i(n)$ &$\langle\dnu\rangle$, $\langle r_{02}\rangle$, $r_{01}(n)$, $r_{10}(n+3)$ & $\nu_i(n)$, $\runo$, $\rdos(n)$\\
Surface correction & BG14 & SC & None & TS15 & BG14 & HK08 & BG14 \\
Optimisation & MCMC & $\chi^2$ minimization & Bayesian & Levenberg-Marquardt & Downhill simplex & $\chi^2$ minimization & Monte Carlo \\
Reference & Appendix~\ref{app:pip} & VSA15 & VSA15 & LG14 & TA15 & TA15 & VSA15 \\
\enddata
\tablecomments{References. Evolutionary models: ASTEC~\citep{ChristensenDalsgaard:2008bi}, Cesam2k~\citep[][]{Morel:2008kw}, GARSTEC~\citep[][]{Weiss:2008jy}, MESA~\citep[][]{Paxton:2011jf}, YREC~\citep[][]{Demarque:2007ij}. Pulsation frequencies: AB94~\citep[][]{Antia:1994vm}, ADIPLS~\citet[][]{ChristensenDalsgaard:2008kr}, LOSC~\citet[][]{Scuflaire:2007fy}. Solar mixture: GN93~\citep{Grevesse:1993vd}, GS98~\citep{Grevesse:1998cy}. Opacities: OPAL96~\citep{Iglesias:1996dp}, OP05~\citep{Badnell:2005ef}, JF05~\citep{Ferguson:2005gn}. Equation of state (EOS): OPAL05~\citep{Rogers:2002cr}. Nuclear reactions: Solar Fusion~\citep{Adelberger:1998iv}, NACRE~\citep{Angulo:1999kp}. Convection theory: MLT58~\citep{BohmVitense:1958vy}, CG68~\citep{1968pss..book.....C}, MLT12~\citep{2012sse..book.....K}, CGM96~\citep{1996ApJ...473..550C}. Diffusion: T94~\citep{Thoul:1994iz}, MP93~\citep{1993ASPC...40..246M}. Surface corrections: HK08~\citep{Kjeldsen:2008kw}, SC~\citep[solar scaled, see appendix A1 in][]{2015MNRAS.452.2127S}, BG14~\citep{Ball:2014gx}, TS15~\citep{Sonoi:2015gy}. Pipelines: VSA15~\citep{2015MNRAS.452.2127S}, LG14~\citep{Lebreton:2014gf}, TA15~\citep{Appourchaux:2015gw}. See text for details.}
\end{splitdeluxetable*}

{\flushleft \ttfamily AIMS}\quad The "Asteroseismic Inference on a Massive Scale" pipeline is described in detail for the first time in this paper and we give a description of it in Appendix~\ref{app:pip}. Briefly, it uses the grid of MESA models from \citet{Coelho:2015ci} and Delaunay tessellation for linear interpolation across the grid. It then applies an MCMC algorithm to find a representative set of models which satisfy the seismic and classic constraints. Stellar properties and uncertainties are obtained by averaging and calculating the standard deviations of the properties from the set of models.

{\flushleft \ttfamily ASTFIT}\quad The "ASTEC FITting" method uses the ASTEC evolutionary code \citep{ChristensenDalsgaard:2008bi} coupled to ADIPLS \citep{ChristensenDalsgaard:2008kr} for computation of theoretical pulsation frequencies in a grid of stellar models. The initial chemical composition is determined exploring a range of values for the galactic enrichment law, and it includes the effects of microscopic diffusion for low-mass stars and three values of the mixing-length efficiency $\alpha_\mathrm{MLT}=1.5, 1.8, 2.1$. The statistical methods applied to determine the stellar properties are described in Appendix~A1 of \citet{2015MNRAS.452.2127S}.

{\flushleft \ttfamily BASTA}\quad The "BAyesian STellar Algorithm" uses precomputed grids of evolutionary models and performs a global search for the optimal solution using the Bayesian approach described by \citet{2015MNRAS.452.2127S}. The current implementation considers microscopic diffusion for masses below 1.2~M$_\odot$, and the NACRE rates with the updated $^{14}N(p,\gamma)^{15}O$ reaction from \citet{Formicola:2004dl}, and overshooting in the exponential decay formulation of \citet{Freytag:1996vw} using the calibrated efficiency determined by \citet{Magic:2010iz}. The original grids have been extended in this work to include lower metallicities as required to match the spectroscopically determined surface composition of some targets.

{\flushleft \ttfamily C2kSMO}\quad The "Cesam2k Stellar Model Optimization" pipeline ({\ttfamily C2kSMO}) uses the Cesam2k evolutionary code coupled to the Li\`{e}ge Oscillations Code and the procedures described by \citet{Lebreton:2014gf}. The input physics considered includes atomic diffusion, convective core overshooting with an efficiency of $d_\mathrm{ov}=0.15\times\mathrm{min}(H_\mathrm{p}, R_\mathrm{conv core})$, convection being treated under the \citet{1996ApJ...473..550C} formalism, and the NACRE rates with the updated $^{14}N(p,\gamma)^{15}O$ reaction from \citet{Imbriani:2005he}. An initial approach to the solution is performed by finding the best fit to the observed values of $\teff$, $\feh$, and $\log\,(g)$ (the latter obtained from the global asteroseismic parameter $\num$), using a chemical enrichment law $\Delta$Y/$\Delta$Z from a solar calibration computed with the same input physics. The resulting age, mass and initial chemical composition are used as starting values in a Levenberg-Marquardt minimisation that adjusts free parameters in the modelling in order to minimise the merit function \citep[see][]{Miglio:2005is}. These free model parameters are the age and mass of the star, the initial helium content and initial $(Z/X)_0$ ratio, and the convective efficiency $\alpha_\mathrm{CGM}$. The observables fitted by this optimisation process are $\feh$, $\teff$, $\log\,(g)$, the lowest observed radial mode, and the frequency separation ratios $\runo$ and $\rdos$. The error bars on the free parameters are obtained as the diagonal coefficients of the inverse of the Hessian matrix, while uncertainties in the other parameters are obtained following Eq.~10 in \citet{Deheuvels:2016ek}.

{\flushleft \ttfamily GOE}\quad The setting of the "GOEttingen" pipeline is very similar to that described by \citet{Appourchaux:2015gw} and \citet{Reese:2016jo}. Since more massive stars are included in the LEGACY sample, the \citet{Ledoux:1947ka} criterion for convection was used with an extra free parameter for the convective overshooting described according to the exponential model of \citet{Freytag:1996vw}. The NACRE thermonuclear reactions are used with the updated rates in $^{14}N(p,\gamma)^{15}O$ from \citet{Imbriani:2005he}. The free adjustable parameters to find the best-fitting model were age, initial metallicity, initial helium, the mixing-length, the overshooting efficiency, and the two terms in the \citet{Ball:2014gx} surface correction.

{\flushleft \ttfamily V\&A}\quad This approach uses a combination of the MESA and ADIPLS codes to compute evolutionary models and pulsation frequencies with the same input physics as listed in Table~3 of \citet{Appourchaux:2015gw}. Diffusion of helium and heavy elements is typically not included for stars with masses greater than 1.35~M$_\odot$, while overshoot is taken into account with an exponentially decaying efficiency for stars in the relevant mass regime (above 1.10~M$_\odot$). For thermonuclear reactions the NACRE compilation is used with the exception of updated rates in the $^{14}N(p,\gamma)^{15}O$ reaction from \citet{Imbriani:2005he}. The properties of stars are determined via $\chi^2$ minimization over a suitable grid of models. This is the only pipeline that uses the large frequency separation as an asteroseismic constraint, obtained by a linear least-squares fit to the observed frequencies as a function of radial order.

{\flushleft \ttfamily YMCM}\quad The "Yale Monte Carlo Method" is applied as described by \citet{2015MNRAS.452.2127S}, with the exceptions of a variable value of the mixing-length parameter and the use of the \citet{Ball:2014gx} formulation to correct for surface effects. The Solar fusion \citep{Adelberger:1998iv} cross section for $^{14}N(p,\gamma)^{15}O$ has been updated according to \citet{Formicola:2004dl}.
\section{Results} \label{s:res}
\subsection{Precision of the stellar properties}\label{ss:pres}
\begin{figure*}[ht!]
\centering
\includegraphics[]{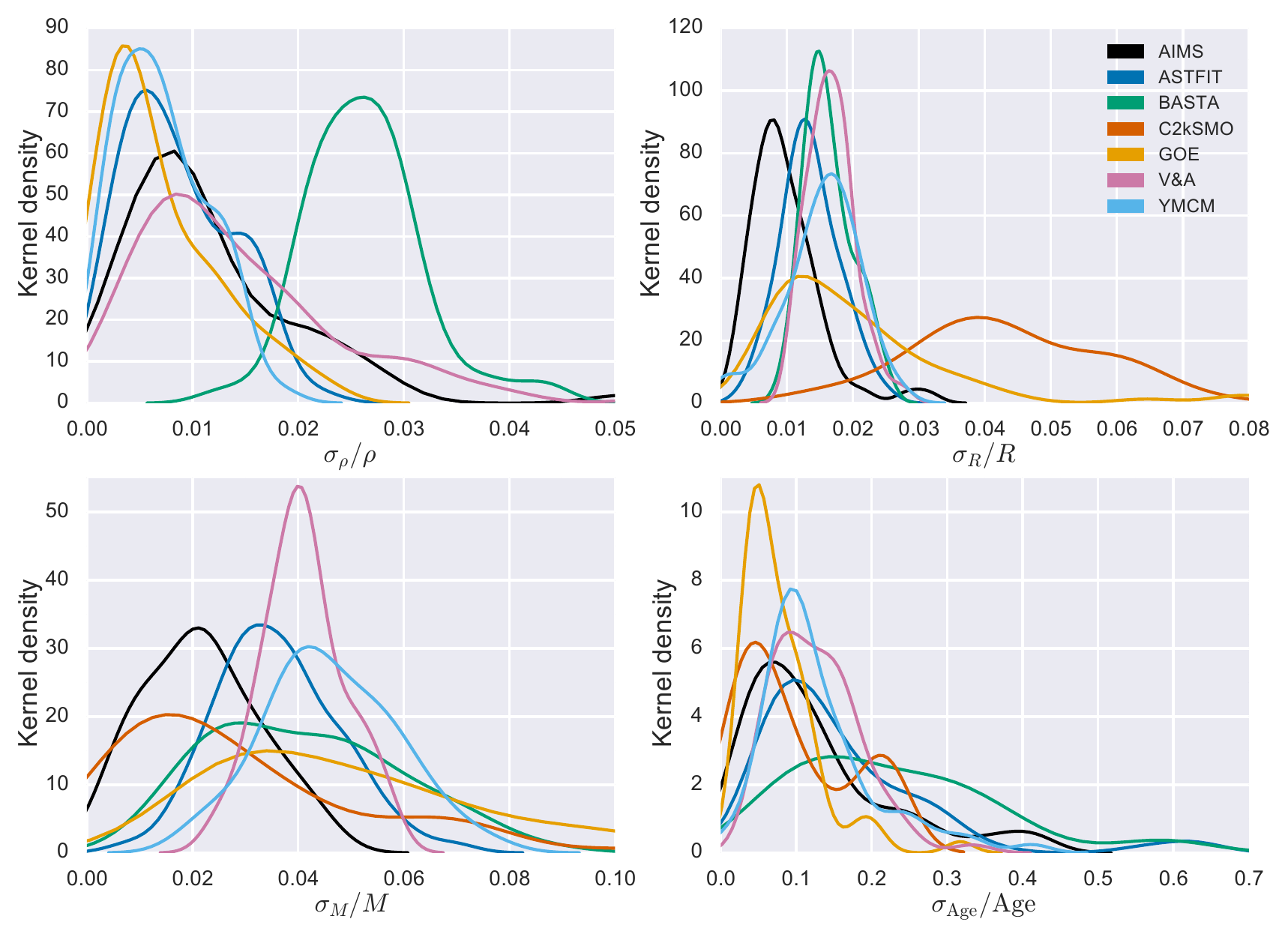}
\caption{Distribution of uncertainties in density, radius, mass, and age for all stars in the sample as determined by each pipeline. When asymmetric error bars are reported these were added in quadrature for the figure. Values for the {\ttfamily C2kSMO} density uncertainties are beyond the scale of the plot and therefore not shown. See text for details. \label{fig:uncert}}
\end{figure*}
We determined stellar properties for the 66 stars in the LEGACY sample using the pipelines described in the previous section. Considering the variety of methods and input physics involved in the process, we choose to report the results from all of pipelines in complete and homogeneous sets given in Appendix~\ref{app:tab}.

Figure~\ref{fig:uncert} shows the uncertainty distributions obtained by each pipeline for the LEGACY sample. Overall, the results agree at the level of the size of their formal uncertainties, with some exceptions for given parameters. In terms of formal errors in density the {\ttfamily BASTA} results are of the order $\sim$2.5\%, a factor of two larger than the majority of pipelines, while uncertainties from {\ttfamily C2kSMO} have median values at the 16\% level and are much larger than those of any other pipeline. The reasons for these discrepancies are related to the details of each individual method: {\ttfamily BASTA} is the only pipeline to fit only frequency ratios, which are less sensitive to the mean density of the star than individual frequencies \citep[see][for a discussion]{2015MNRAS.452.2127S}. On the other hand, the local minimisation Levenberg-Marquardt algorithm used by {\ttfamily C2kSMO} does not determine uncertainties for density so these are derived by (uncorrelated) error propagation from mass and radius, which results in much higher formal errors than all other methods.

In Fig.~\ref{fig:uncert}, the results for radius and mass also reveal some differences between the pipelines: {\ttfamily C2kSMO} has much larger radius uncertainties and formal mass errors only slightly smaller than the rest of the methods, resulting in the large density fractional uncertainties. Age distributions show some pipelines with fractional errors below 10\% ({\ttfamily C2kSMO} and {\ttfamily GOE}) while the rest encompass values between 10-30\%, which is in much better agreement with the level of age uncertainties determined from hare-and-hounds exercises \citep{Reese:2016jo} and those normally obtained from individual frequency fitting of {\it Kepler} main-sequence targets \citep[e.g.,][]{Mathur:2012bj,SilvaAguirre:2013in,Lund:2014ez,Lebreton:2014gf,2015MNRAS.452.2127S}.
\begin{figure}[ht!]
\centering
\includegraphics[width=\linewidth]{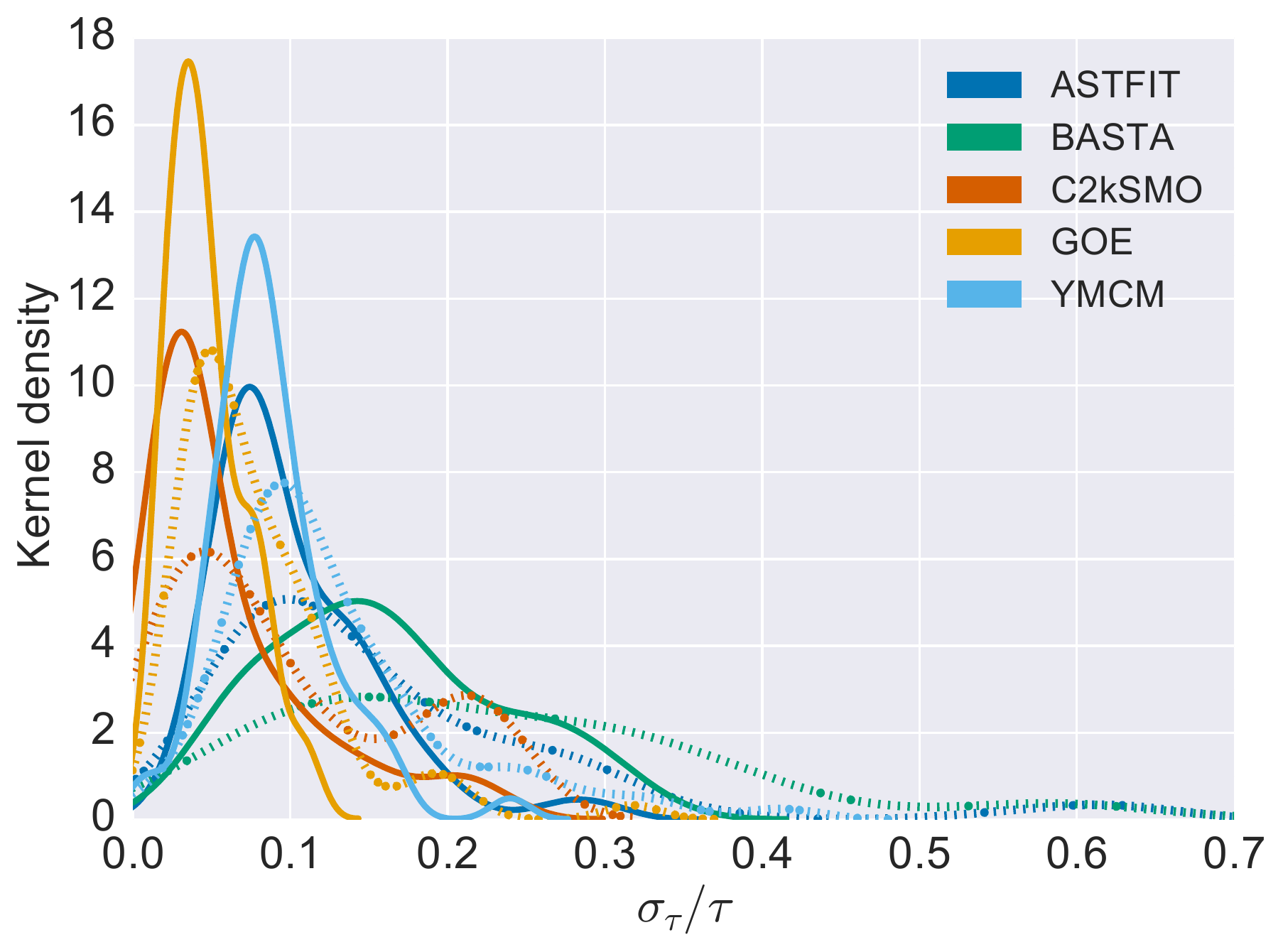}
\caption{Age uncertainty distributions normalised to the TAMS age (solid) or the seismically determined age (dotted) of the target. See text for details. \label{fig:tams}}
\end{figure}

The distributions in age show a tail towards large fractional error values (up to $\sim$60\%), much higher than the expected formal uncertainties from asteroseismic determinations. These are dominated by the youngest stars in our sample that have statistical errors of similar magnitude than the oldest stars. A more physically motivated scale to compare the age uncertainty distributions is given by the Terminal Age Main Sequence (TAMS), which we define as the point where the remaining central hydrogen content in a stellar model reaches $1\times10^{-5}$. Five pipelines were able to produce a proxy for the TAMS values of our targets by evolving their best-fitting model beyond the main-sequence phase. Figure~\ref{fig:tams} shows the uncertainty distributions normalised by current and TAMS age, where the tail at large fractional uncertainties is clearly reduced by the change in scale while the peak of the distribution remains very close to its original value. The latter is a consequence of an asteroseismic detection bias which favours more evolved stars, as the oscillation amplitudes scale proportionally with stellar luminosity \citep[e.g.,][]{Kjeldsen:1995tr}.

In summary, stellar properties of our targets are determined to varying degrees of precision by the pipelines, with median uncertainties ranging from 0.5\%$-$2.6\% in density (excluding the {\ttfamily C2kSMO} results, see above); 1.3\%$-$4.2\% in radius; 2.3\%$-$4.5\% in mass; and 6.7\%$-$20\% in age. This level of precision is comparable to previous asteroseismic studies of smaller samples using individual frequencies of oscillations, and makes this set the best-characterised main-sequence sample from the nominal {\it Kepler} mission.
\subsection{Comparison between pipelines}\label{ss:comp}
\begin{figure*}[ht!]
\centering
\includegraphics[]{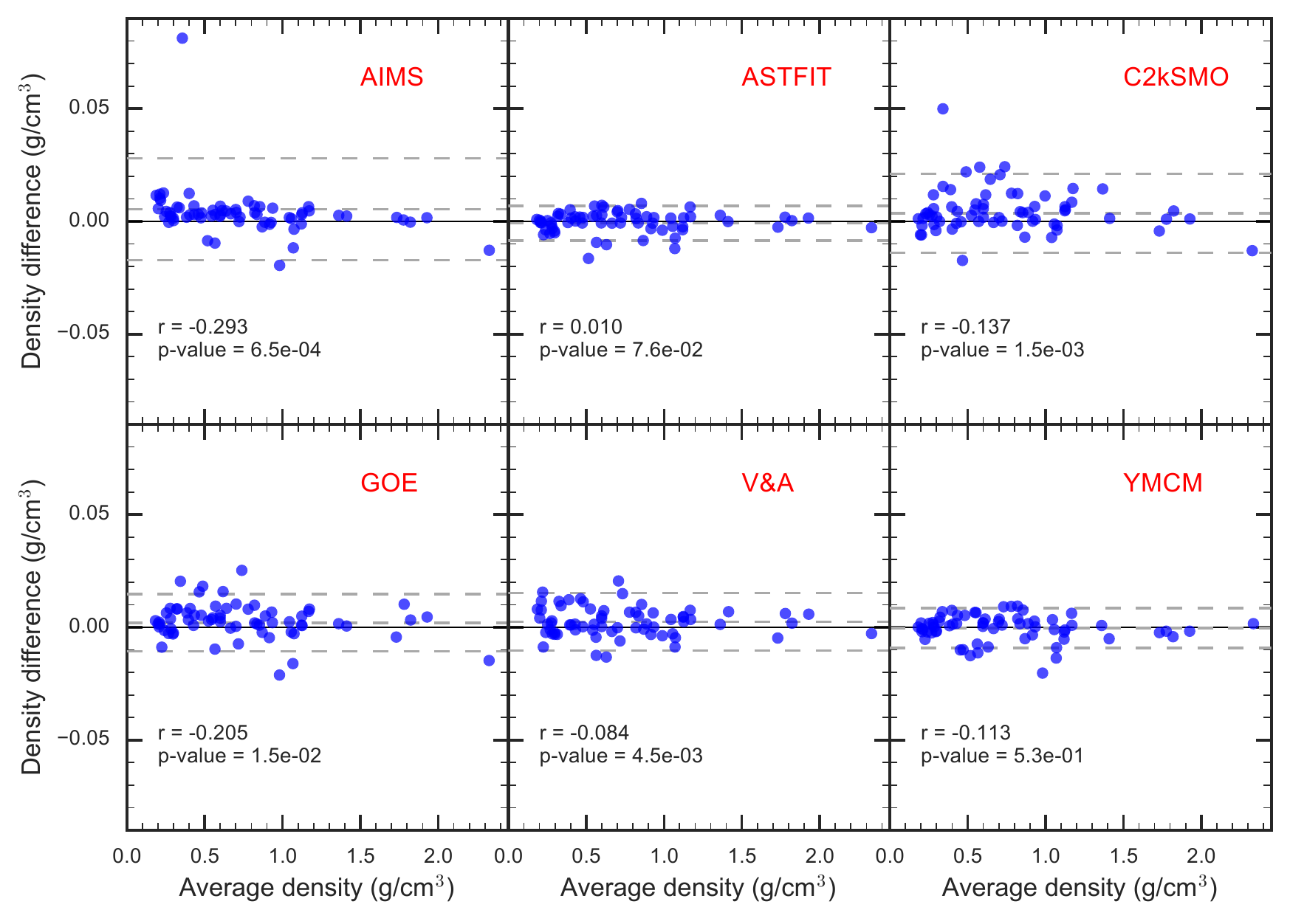}
\caption{Densities determined from each pipeline compared to those obtained with {\ttfamily BASTA}. The abscissa shows the average density value calculated between the pipeline and {\ttfamily BASTA} while the ordinate gives the density differences in the sense (pipeline$-${\ttfamily BASTA}). Solid lines mark the zero level to guide the eye while dashed lines correspond to the mean and twice the standard deviation of the differences weighted by their uncertainties (added in quadrature). Pearson correlation coefficients $r$ and p-values are also given. See text for details.\label{fig:dens}}
\end{figure*}

\begin{figure*}[ht!]
\centering
\includegraphics[]{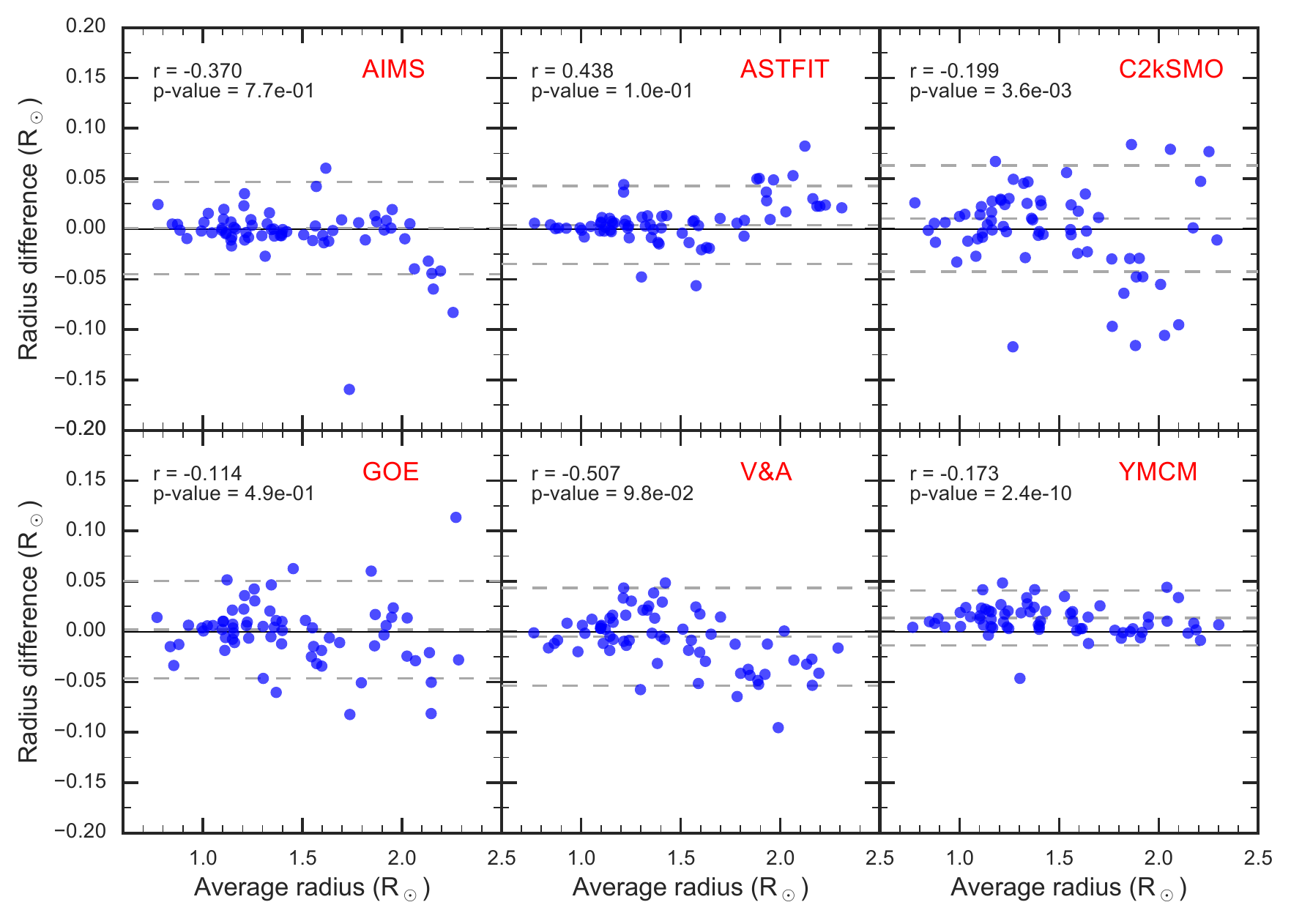}
\caption{Same as figure~\ref{fig:dens} but for radius determinations.\label{fig:rad}}
\end{figure*}

\begin{figure*}[ht!]
\centering
\includegraphics[]{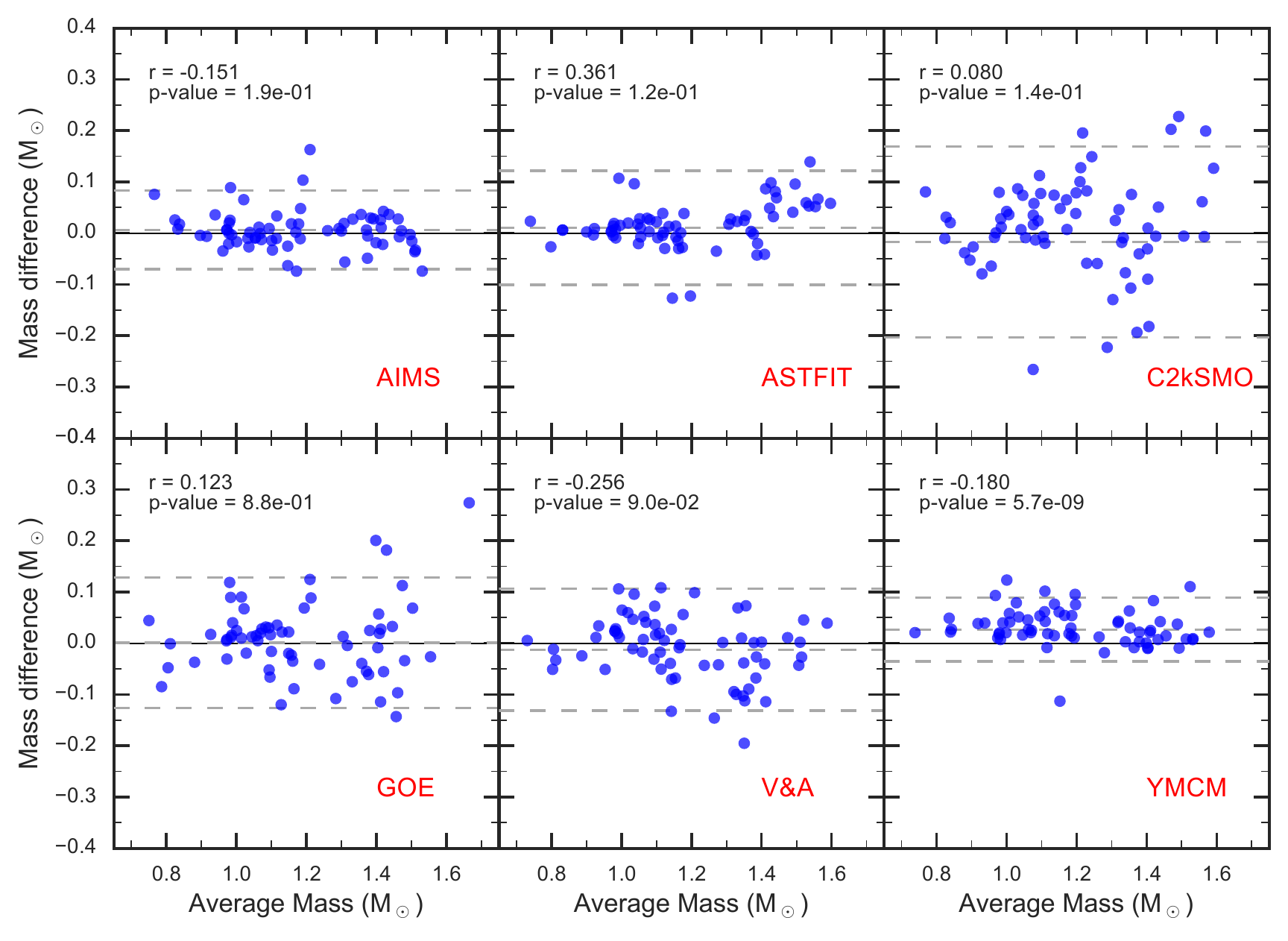}
\caption{Same as figure~\ref{fig:dens} but for mass determinations.\label{fig:mass}}
\end{figure*}

\begin{figure*}[ht!]
\centering
\includegraphics[]{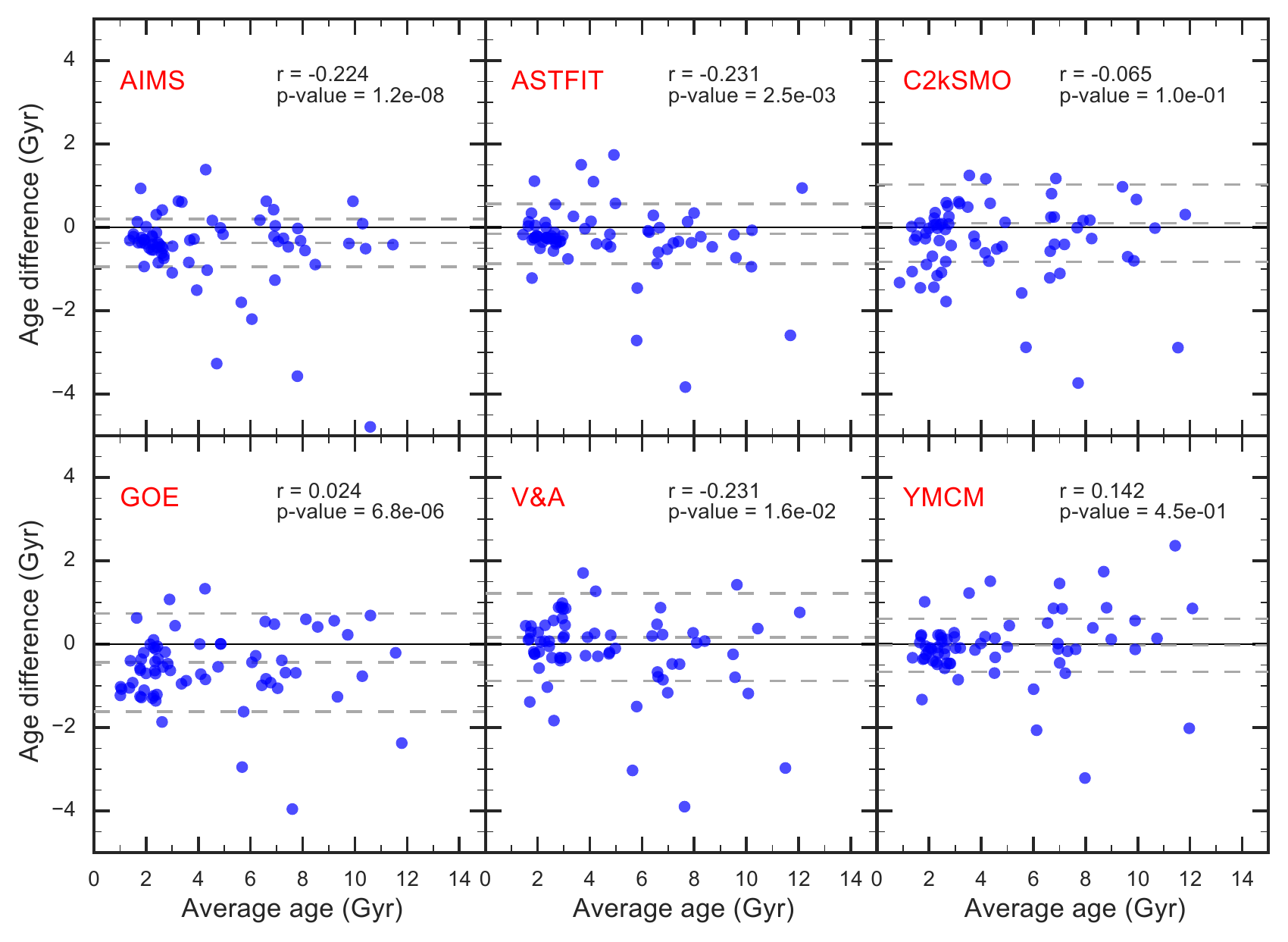}
\caption{Same as figure~\ref{fig:dens} but for age determinations.\label{fig:age}}
\end{figure*}

The stellar properties predicted by each pipeline represent an estimate of the true physical parameters of our sample based on a given set of data. For the large majority of our targets, we lack an independent verification of these stellar properties that would allow us to test the validity of our results (but see Section~\ref{s:checks} below for some exceptions). It is customary in these cases to compare the results across methods by selecting a reference pipeline and plot the fractional differences in a given quantity as a function of the result predicted by the reference. Since {\it a priori} we cannot consider any set of results to be an accurate representation of the underlying stellar properties of the sample, this approach can introduce spurious correlations misleading the analysis \citep[see, e.g.,][]{Ludbrook:1997vd}. An alternative is to use the Altman-Blandt method for analysing differences \citep{Bland:1986wu,Bland:1995vm}, consisting in calculating the actual difference between the results of a pipeline and the reference, and plotting them against the corresponding mean values. Figures~\ref{fig:dens},~\ref{fig:rad},~\ref{fig:mass}, and ~\ref{fig:age} depict these differences for density, radius, mass, and age using the {\ttfamily BASTA} results as a reference. In each panel we include the results of computing the Pearson product-moment correlation coefficient $r$ of the mean values, as well as the p-value of a one-sample $t-$test of the weighted mean of the differences.

If there is no proportional bias between the results of a pipeline and the reference, the regression of the difference on means should have a slope of zero. For a sample of the size considered here, it translates into an absolute value of $\left| r \right|>0.25$ for the bias to be significant at the 5\% level. Except for the {\ttfamily AIMS} comparison, $r$ values in Fig.~\ref{fig:dens} reveal no proportional bias in density as a function of its mean value, a reassuring result considering that the large frequency separation is reproduced by all methods when fitting individual frequencies and it scales to good approximation with the square root of the mean stellar density (see Eq.~\ref{eq:dnuscal}).

We include in the comparison plots dashed lines showing the mean and twice the standard deviation of the differences (weighted by the uncertainties). If there were no fixed bias between the results then the mean of the differences should be zero. To test this, we performed a one-sample $t-$test of the weighted mean of the difference in each stellar property, where resulting p-values below 0.05 reject the H$_0$ hypothesis of zero mean differences and reveal a fixed bias between the methods. Thus, we find no such biases in the comparisons with {\ttfamily ASTFIT}, and {\ttfamily YMCM}, while the other four pipelines show a statistically significant (and positive) fixed bias.

To gain further understanding of these results we followed up with a similar analysis of the results for radius and mass, showed in Figs.~\ref{fig:rad} and~\ref{fig:mass}. There is a clear proportional bias in radius between {\ttfamily BASTA} and {\ttfamily ASTFIT} (increasing), and between {\ttfamily BASTA} and {\ttfamily V\&A} (decreasing). Considering that this correlation was not observed in any pipeline for the density results, it must be compensated in the mass determinations as confirmed by Fig.~\ref{fig:mass} where the same two pipelines show a proportional bias according to their $r$ values. On the other hand, the proportional bias in density between {\ttfamily BASTA} and {\ttfamily AIMS} is a result of a similar bias in radius, as the mass comparison between these pipelines shows no traces of such bias.

The results of the $t$-test for radius revealed a fixed bias for {\ttfamily C2kSMO} and {\ttfamily YMCM}, both cases predicting larger radii than {\ttfamily BASTA}. In the case of {\ttfamily C2kSMO} a similar behaviour was already present in the density: since this pipeline extracts this quantity from propagation of the mass and radius determinations we expected one of these quantities to be the underlying effect causing the bias. The mass comparison does not show a fixed bias between {\ttfamily C2kSMO} and {\ttfamily BASTA}.

Masses determined with {\ttfamily YMCM} are the only ones showing a statistically significant fixed bias (higher values), while the {\ttfamily ASTFIT} and {\ttfamily V\&A} results hint towards proportional biases in mass (cf, Fig.~\ref{fig:mass}). One would naively expect a similar behaviour in the age determination, since mass is the main property controlling the main-sequence lifetime of a star. Figure~\ref{fig:age} reveals that there are no correlations as a function of age for any of the pipeline results, and also that the fixed bias between {\ttfamily BASTA} and {\ttfamily YMCM} is not significant. However, only {\ttfamily C2kSMO} and {\ttfamily YMCM} do not show a significant fixed bias while all other pipelines do predict such a difference. The lack of correspondence between masses and ages is a result of the differences in evolutionary and pulsation codes employed by the methods, as well as the included input physics \citep[see Table~\ref{tab:pipelines}, also the discussion in Sect.~4 of][]{2015MNRAS.452.2127S}.

In summary, the results from different pipelines show some fixed and proportional biases for given stellar properties, but overall there is an excellent level of agreement considering the variety of evolutionary and pulsation codes, input physics employed, as well as the differences in the seismic quantities fitted by each pipeline. This heterogeneity across methods as seen from Table~\ref{tab:pipelines} makes it difficult to isolate the unique physical culprit of a particular outlier in our sample. Thus, we have focused our comparison on general trends with description of individual cases when possible, and note that detailed discussions of the impact of changing the input physics on asteroseismically derived properties are given by \citet{Lebreton:2014gf} and \citet{2015MNRAS.452.2127S}.

\subsection{Initial helium abundance and mixing-length parameter}\label{ss:trends}
Four of the pipelines used in our analysis include initial abundances and efficiency of convection as free parameters that are determined during the optimisation process. We can therefore compare the resulting values with expectations based on chemical enrichment of the Galaxy and predictions of 3D hydrodynamical simulation of atmospheres.

One of the most poorly constrained ingredients in modelling solar-type stars is the helium content, as its abundance cannot be determined from spectroscopic observations. The transformation from the observed metallicity $\feh$ to fractional abundances $X, Y, Z$ is achieved by assuming a relation between heavy elements and helium content known as the chemical enrichment law $\Delta Y / \Delta Z$. The existence and value of such a relation have long been debated, and current understanding favours the range between $1 \leq \Delta Y / \Delta Z \leq 3$ as obtained from different sets of indicators \citep[see e.g.,][and references therein]{1999A&A...350..587L,Jimenez:2003il,2006AJ....132.2326B,Casagrande:2007ck,Serenelli:2010gu}. The relation is anchored to a known point in the helium-heavy elements plane, normally the values obtained from Standard Big Bang Nucleosynthesis (SBBN) calculations $Z_0=0$ and $Y_0=0.248$ \citep{Steigman:2010gz}.

\begin{figure*}[ht!]
\centering
\includegraphics[]{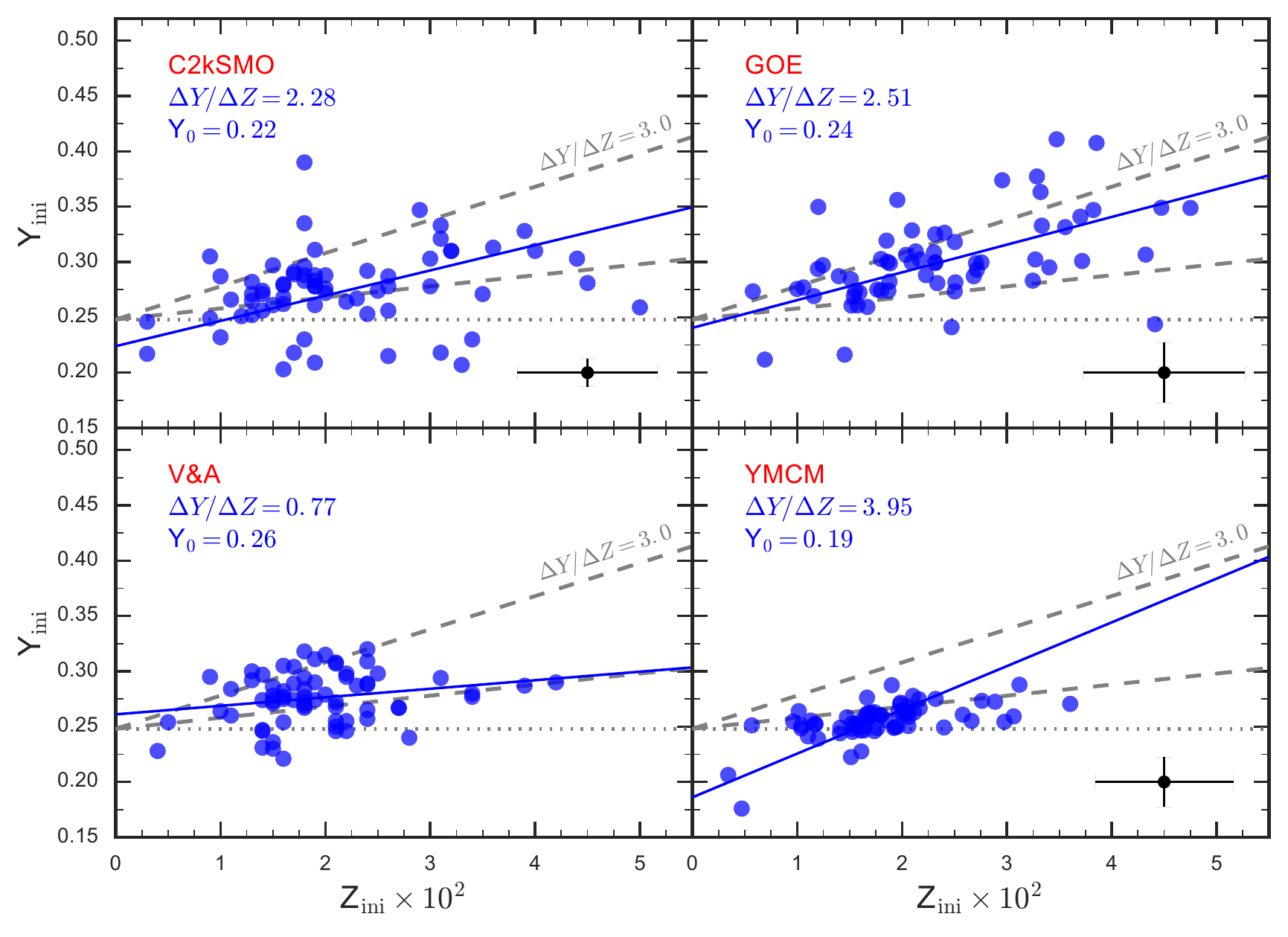}
\caption{Initial helium abundance as a function of initial heavy elements abundance for the pipelines that do not constrain the chemical composition with a galactic enrichment law. Two dashed grey lines depict different slopes in the $\Delta Y / \Delta Z$ relation: 3.0 (upper), 1.0 (lower). The dotted grey line shows the primordial helium abundance predicted by SBBN \citep{Steigman:2010gz}, while the solid blue lines present a linear least-square fit to the results from each pipeline, characterised by the slope and intercept given in each panel. Median uncertainties are plotted to avoid clutter (black circles, except for V\&A that does not provide them for these parameters). See text for details.\label{fig:dydz}}
\end{figure*}

Figure~\ref{fig:dydz} shows the initial helium and heavy element abundance for all targets as predicted by the four pipelines freely varying the composition as part of the optimisation. All panels show cases where $Y_\mathrm{ini}$ is below the measurement from SBBN, a problem commonly appearing in asteroseismic results fitting individual oscillation frequencies \citep[see e.g.,][]{Mathur:2012bj,Metcalfe:2014ig,2015MNRAS.452.2127S}.

A linear fit to the results from each pipeline (weighted by the uncertainties when available) shows slopes compatible with the broad range of $1 \leq \Delta Y / \Delta Z \leq 3$ for the {\ttfamily C2kSMO} and {\ttfamily GOE} pipelines, while the fit to the {\ttfamily V\&A} predictions returns a slightly lower slope. The {\ttfamily YMCM} abundances cluster systematically below the $\Delta Y / \Delta Z = 1$ line and in combination with sub-SBBN helium for its most metal-poor stars predict the highest slope of all targets.

Despite the aforementioned differences, the overall asteroseismic predictions are broadly consistent with the slopes commonly adopted as galactic chemical enrichment laws. The recent measurement of surface helium abundance using asteroseismology in the binary 16 Cyg A\&B \citep{Verma:2014fx} and the detection of glitches associated with the HeII ionisation zone in all stars from our sample (Verma et al. 2016, submitted) open the possibility for detailed studies that can potentially put firm constraints on $Y_\mathrm{ini}$, and it will be pursued in a future study.

Another poorly constrained parameter that has a huge impact in the modelling results of solar-type stars is the efficiency of convection, parameterised in both the mixing-length \citep{BohmVitense:1958vy} and the full spectrum of turbulent eddies \citep{1996ApJ...473..550C} theories. Normally this efficiency is calibrated to the Sun and used across the HRD, but asteroseismic fitting allows us to obtain a determination of this parameter by including it in the optimisation process. These predictions have not been tested to date due to the lack of benchmarks for comparison, but a new approach is now possible by extracting information about convective efficiency from large grids of 3D hydrodynamical simulations of stellar atmospheres. These simulations follow the properties of convection in the outer stellar layers without the need of free parameters, and can be used to calibrate the expected mixing-length parameter by matching a 1D envelope model to a suitable average of the 3D simulations \citep[see][for a description of the procedure]{Trampedach:2014fo, 2015A&A...573A..89M}.
\begin{figure*}[ht!]
\centering
\includegraphics[]{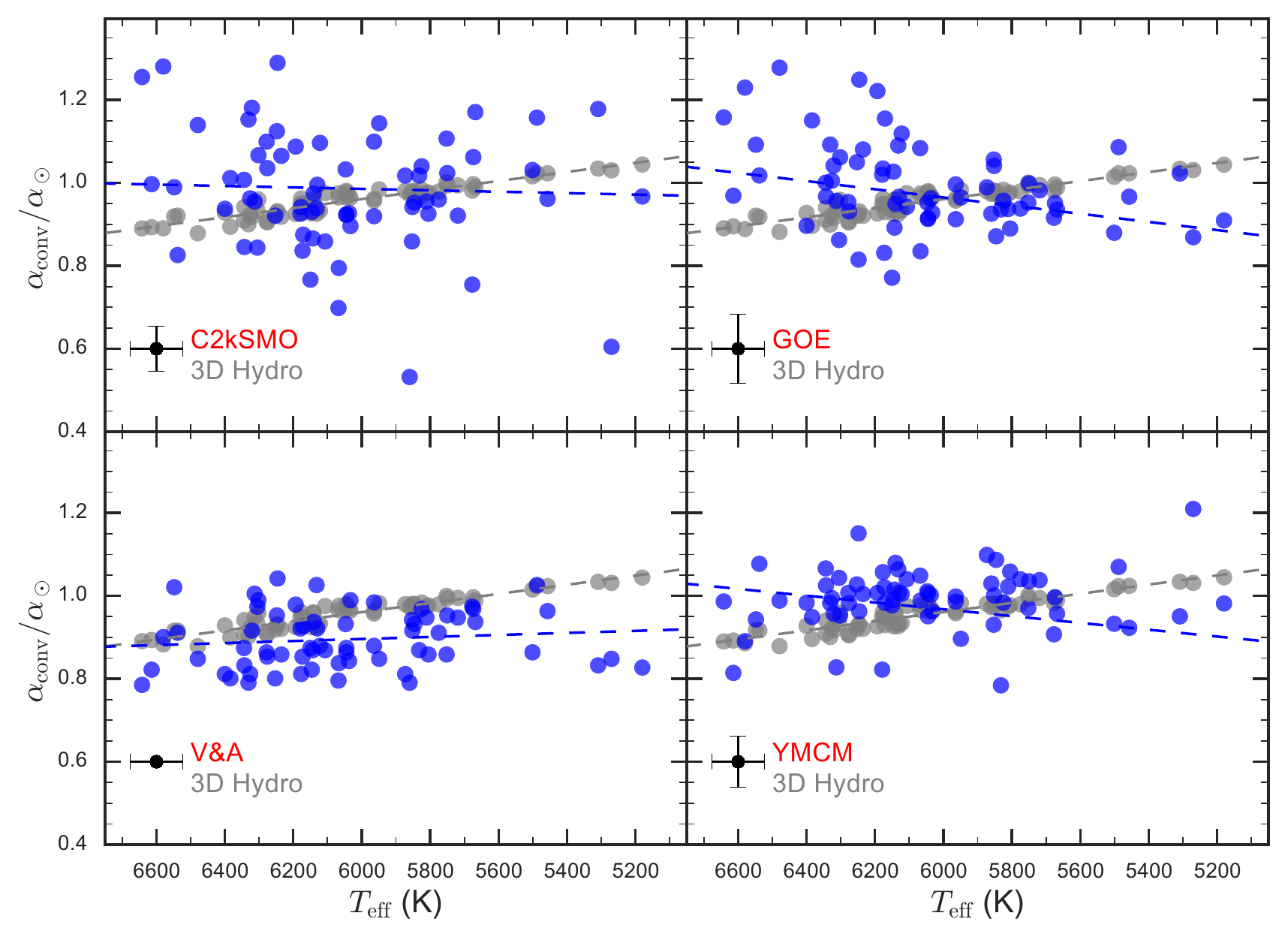}
\caption{Mixing-length parameter as a function of effective temperature for the LEGACY sample. Plotted on the Y-axis is the ratio between the $\alpha$ parameter of the targets and the solar mixing length determined for each pipeline (blue circles) or from 3D hydrodynamical simulations (grey circles). Dashed lines represent a linear fit weighted by the uncertainties when available. To avoid clutter, median uncertainties from each pipeline are plotted (black circles). The {\ttfamily C2kSMO} results are computed with the \citet{1996ApJ...473..550C} formalism of convection. See text for details.\label{fig:alphaconv}}
\end{figure*}

In particular, the Stagger collaboration has produced a grid of more than 200 simulations covering a wide range in metallicity, effective temperature and surface gravity. An equivalent mixing-length parameter based on these 3D simulations is available as a function of these properties \citep{2015A&A...573A..89M}, and we make a first comparison of their results to our determinations in Fig.~\ref{fig:alphaconv}. For a given pipeline we show, as a function of $\teff$, the ratio between the mixing-length parameter of each star and the value for the Sun obtained from a solar calibration. In the 3D hydrodynamical case, the solar value is determined from an average solar simulation. The rationale behind presenting the comparison in terms of ratios in the convective efficiency is to minimise as much as possible the effects of the different input physics used in each pipeline and the atmospheric simulations. We caution that the comparison with {\ttfamily C2kSMO} is shown only for guidance, as this pipelines uses the \citet{1996ApJ...473..550C} theory for convection whose efficiency value is not directly comparable to the mixing-length one.

None of the pipelines in our study follow the decreasing trend in convective efficiency as a function of $\teff$ predicted by the simulations. In fact, two of the pipelines are consistent with a relatively flat relation ({\ttfamily C2kSMO} and {\ttfamily V\&A}), while the {\ttfamily GOE} and {\ttfamily YMCM} pipelines seem to suggest the opposite behaviour as the observed in the 3D case. It is worth noticing that {\ttfamily V\&A} predicts mixing-length parameters systematically lower than the solar one. The results from 3D simulations in terms of an HRD-position-dependent convective efficiency are starting to be adopted by the stellar community in evolutionary calculations \citep[see, e.g.,][for the first examples]{Salaris:2015be,Mosumgaard:2016vl}, and our sample of stars would be perfectly suited to test the impact of including results from 3D simulations on asteroseismically derived properties.
\section{Accuracy of stellar properties} \label{s:checks}
In the previous section we compared the results from the different pipelines assuming that the real values of the predicted stellar properties were not known, and therefore each determination of a given quantity was treated as an independent method of measurement. While this is true for the large majority of our targets, in some cases additional constraints on the stellar properties are available from interferometry, parallaxes, or binarity. The following sections put our derived quantities under scrutiny using these additional data.
\subsection{The Sun}\label{ss:sun}
\begin{figure}[ht!]
\centering
\includegraphics[width=\linewidth]{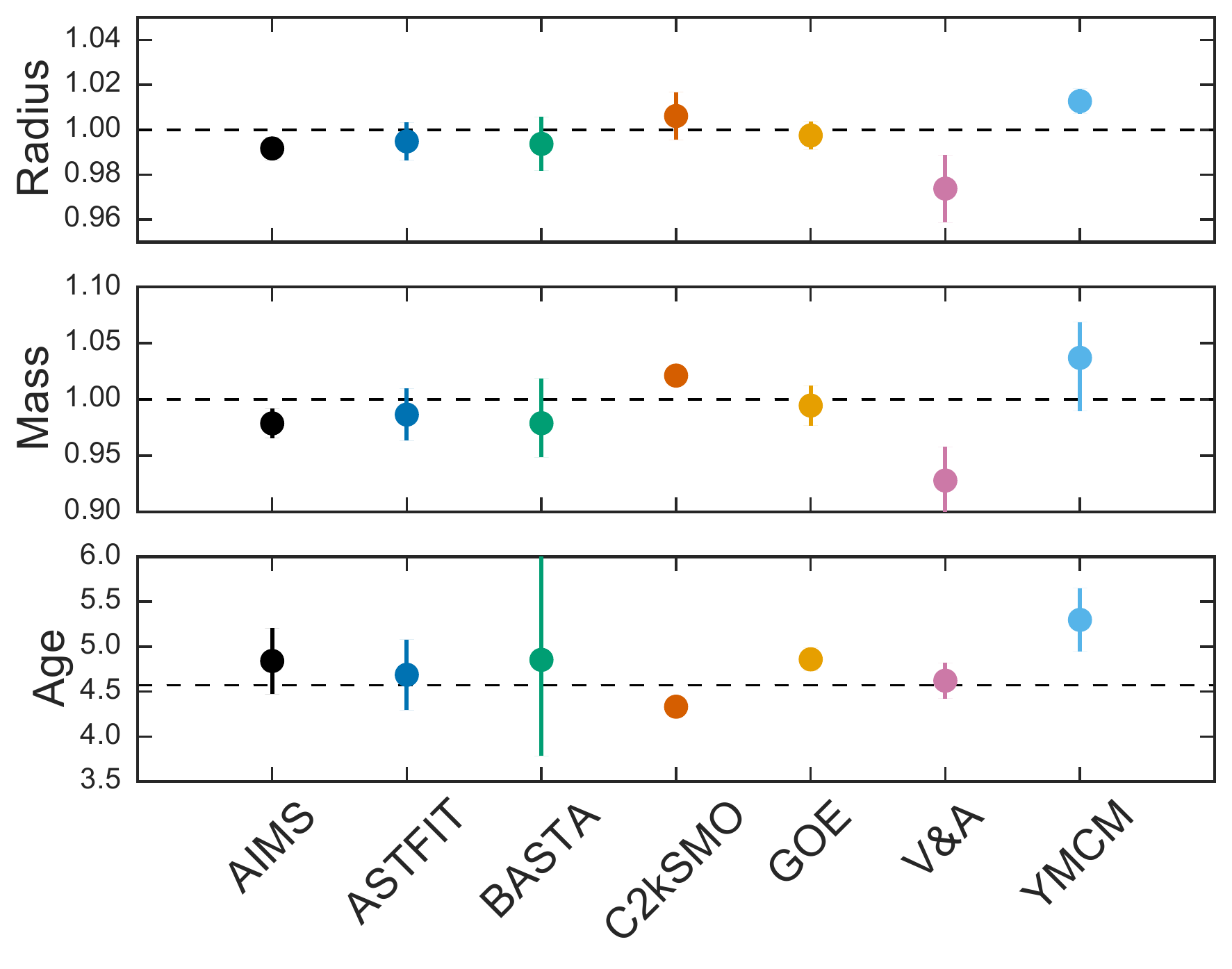}
\caption{The radius, mass, and age of the Sun determined by the pipelines for the concealed solar dataset. See text for details.\label{fig:sunprop}}
\end{figure}
Our own sun is the star we have best characterised in the Universe and displays a rich spectrum of stochastically-excited modes of oscillations. To test the accuracy of our pipelines we included in the target list circulated among all modellers a set of frequencies determined from solar data degraded in quality to match that of the {\it Kepler} mission (see Lund et al. 2016, submitted, for details). The true identity of this star was concealed from the modellers, and the results presented here include the same relevant input physics as given in Table~\ref{tab:pipelines} and fitting methods described in Sec~\ref{s:methods}. Besides the frequencies, we provided a temperature of $\teff=5752\pm77$~K and a surface composition of $\feh=-0.02\pm0.15$~dex determined from a random realisation of Gaussian noise perturbations of the solar values and having uncertainties representative of the rest of the {\it Kepler} LEGACY sample. The results are given in Table~\ref{tab:sun} and shown in Fig.~\ref{fig:sunprop}, showing that in general most methods return values compatible with the solar properties. However, some discrepant results appear that deserve further discussion.

The mass reported by {\ttfamily C2kSMO} are more than 3$\sigma$ away from the Sun despite determining values comparable to other pipelines that agree with solar properties within one standard deviation. The reason is the smaller uncertainties in this parameter returned by this method, a feature which was already visible in relation to Fig.~\ref{fig:uncert}. A similar issue is found in the derived mass and radius from {\ttfamily AIMS}, where the results lie around 2$\sigma$ away from the true values. It is interesting to see the impact of diffusion on the solar age: while two methods ({\ttfamily C2kSMO} and {\ttfamily YMCM}) predict central masses higher than 1~M$_\odot$, the age of the latter is not lower than the solar one due to the lack of microscopic diffusion in the modelling (see Table~\ref{tab:pipelines}). A test including this effect in the {\ttfamily YMCM} modelling results in an estimated age of $\tau=4.83$~Gyr, in better agreement with the solar value of 4.57~Gyr.

The solar properties determined by the {\ttfamily V\&A} method favour a sub-solar mass and a surface helium abundance higher than measured by helioseismology (see Table~\ref{tab:sun}). This anti-correlation between mass and helium is expected from homology relations \citep[e.g.,][]{Weiss:2005vv} and has been reported before in asteroseismic modelling methods using local optimisation (e.g, \citet{Mathur:2012bj,Metcalfe:2014ig}, see also Section~4.3 in \citet{Lebreton:2014gf} and Section~4.4 in \citet{2015MNRAS.452.2127S}).

According to homology relations there is a strong dependence of the stellar luminosity on the mean molecular weight. Since the luminosity is related to the effective temperature and radius, once these are determined there is a narrow range of combinations of mass and helium abundance that can comply with this constraint. The {\ttfamily V\&A} results for the Sun reveal a bimodal distribution in mass, and the most probably of these peaks is reported in Table~\ref{tab:sun}. However, the secondary solution is in better agreement with the solar properties, yielding a mass of 1.036~M$_\odot$ and surface helium of Y$_\mathrm{sup}=0.237$. We note that when comparing the two solutions, the secondary one cannot be ruled out at the 2$\sigma$ level, and analysis of the amplitude of frequency glitches in the spectrum as done by \citet{Verma:2014fx,Verma:2014jv} favours the high-mass results. This type of study can help breaking the degeneracy between mass and helium abundance and will be considered it a subsequent paper.

Finally we turn to parameters that have been successfully determined from helioseismic inferences, namely the depth of the convective envelope and the helium surface abundance \citep[see][for a review]{ChristensenDalsgaard:2002gr}. In both cases, all but the {\ttfamily V\&A} results give agreement within 1$\sigma$ to the helioseismic values, giving us confidence about the stellar parameters determined from these techniques for stars like the Sun.
\begin{deluxetable*}{llllllll}
\tablecaption{Solar properties determined from each pipeline. See text for details. \label{tab:sun}}
\tablewidth{0pt}
\tablehead{
\colhead{}& \colhead{Mass (M$_\odot$)} &  \colhead{Radius (R$_\odot$)} &  \colhead{Age (Myr)}  &  \colhead{Luminosity (L$_\odot$)} &  \colhead{Density (g/cm$^3$)} &  \colhead{Y$_\mathrm{sup}$} &  \colhead{R$_\mathrm{BCZ}$ (R$_\odot$)}  
}
\startdata
{\ttfamily AIMS}&$ 0.979\pm 0.013$&$ 0.992\pm 0.005$&$ 4840\pm 367$&$ 1.060\pm 0.048$&$1.413\pm 0.004$&--&--\\
{\ttfamily ASTFIT}&$ 0.986\pm 0.023$&$ 0.994\pm 0.008$&$ 4686\pm 393$&$ 0.972\pm 0.052$&$1.411\pm 0.003$&$0.249\pm 0.009$&$0.71\pm 0.007$\\
{\ttfamily BASTA}&$ 0.978^{+0.039}_{-0.030}$&$ 0.993^{+0.012}_{-0.012}$&$ 4852^{+1181}_{-1069}$&$ 0.976^{+0.054}_{-0.052}$&$1.411^{+0.021}_{-0.022}$&$0.247^{+0.012}_{-0.01}$&$0.713^{+0.009}_{-0.009}$\\
{\ttfamily C2kSMO}&$ 1.021\pm 0.003$&$ 1.006\pm 0.010$&$ 4331\pm 85.$&$ 1.084\pm 0.048$&$1.412\pm 0.048$&$0.245\pm 0.003$&$0.715\pm 0.004$\\
{\ttfamily GOE}&$ 0.997\pm 0.006$&$ 0.995\pm 0.018$&$ 4859\pm 128$&$ 0.947\pm 0.041$&$1.412\pm 0.002$&$0.234\pm 0.009$&$0.720\pm 0.003$\\
{\ttfamily V\&A}&$ 0.927\pm 0.030$&$ 0.973\pm 0.015$&$ 4621\pm 200$&$ 0.937 $&$1.418\pm 0.006$&$0.277$&$0.725$\\
{\ttfamily YMCM}&$ 1.037^{+0.031}_{-0.047}$&$ 1.012^{+0.005}_{-0.005}$&$ 5297^{+350}_{-350}$&$ 1.008^{+0.043}_{-0.042}$&$1.406^{+0.001}_{-0.001}$&$0.248^{+0.01}_{-0.01}$&$0.716^{+0.003}_{-0.003}$\\
\enddata
\tablecomments{The surface helium abundance (Y$_\mathrm{sup}$) and radius of the base of the convective envelope (R$_\mathrm{BCZ}$) are not determined by {\ttfamily AIMS}, while {\ttfamily V\&A} reports only values from the best fit model (and thus no uncertainties) for these parameters and the luminosity. For reference, the solar age is determined to be 4.57~Gyr \citep[][]{1995RvMP...67..781B} while helioseismology predicts Y$_\mathrm{sup}=0.248\pm0.003$ \citep{Basu:1998cq} and R$_\mathrm{BCZ}=0.713\pm0.001$~R$_\odot$ \citep{ChristensenDalsgaard:1991iv}}
\end{deluxetable*}

\subsection{Interferometry and photometry}\label{ss:int}
Independent radius determinations provide another check on the stellar properties derived from asteroseismology. Results from interferometric campaigns with CHARA combined with Hipparcos parallaxes have started to appear for targets where oscillations have been detected by {\it Kepler}, but currently amount to a handful in the main-sequence and red giant phases \citep{Huber:2012iv,White:2013bu,Johnson:2014id}. In fact, only five of our targets have published interferometric radii that we can compare to the determinations from asteroseismology.

\begin{figure}[ht!]
\centering
\includegraphics[width=\linewidth]{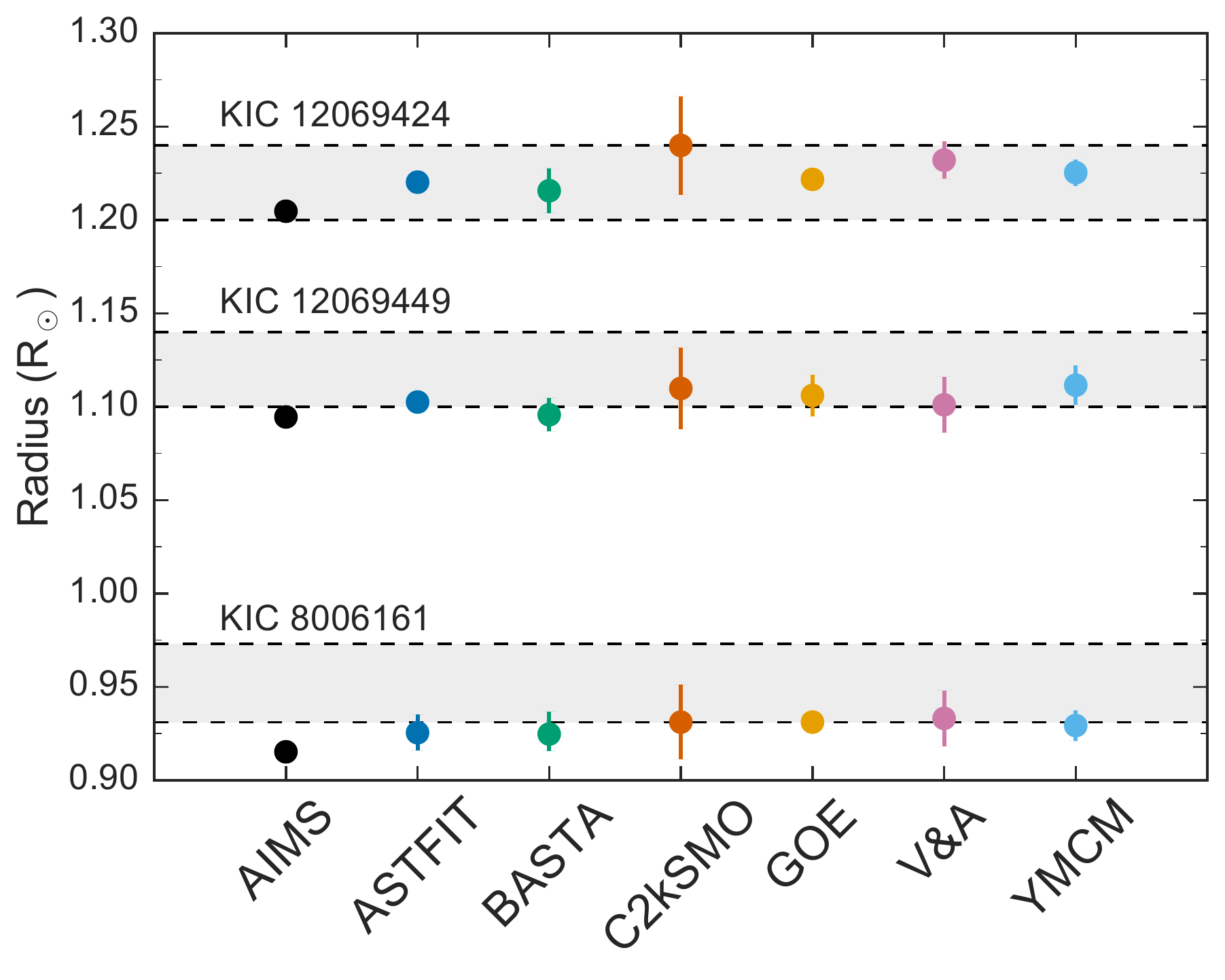}
\caption{Comparison of interferometric radii from \citet{Huber:2012iv} (KIC~8006161) and \citet{White:2013bu} (16~Cyg A\&B, KIC~12069424 and KIC~12069449) with those determined seismically by each pipeline. Grey shaded areas and horizontal dashed lines depict the standard deviation reported in the interferometric measurements.\label{fig:int}}
\end{figure}

\begin{figure}[ht!]
\centering
\includegraphics[width=\linewidth]{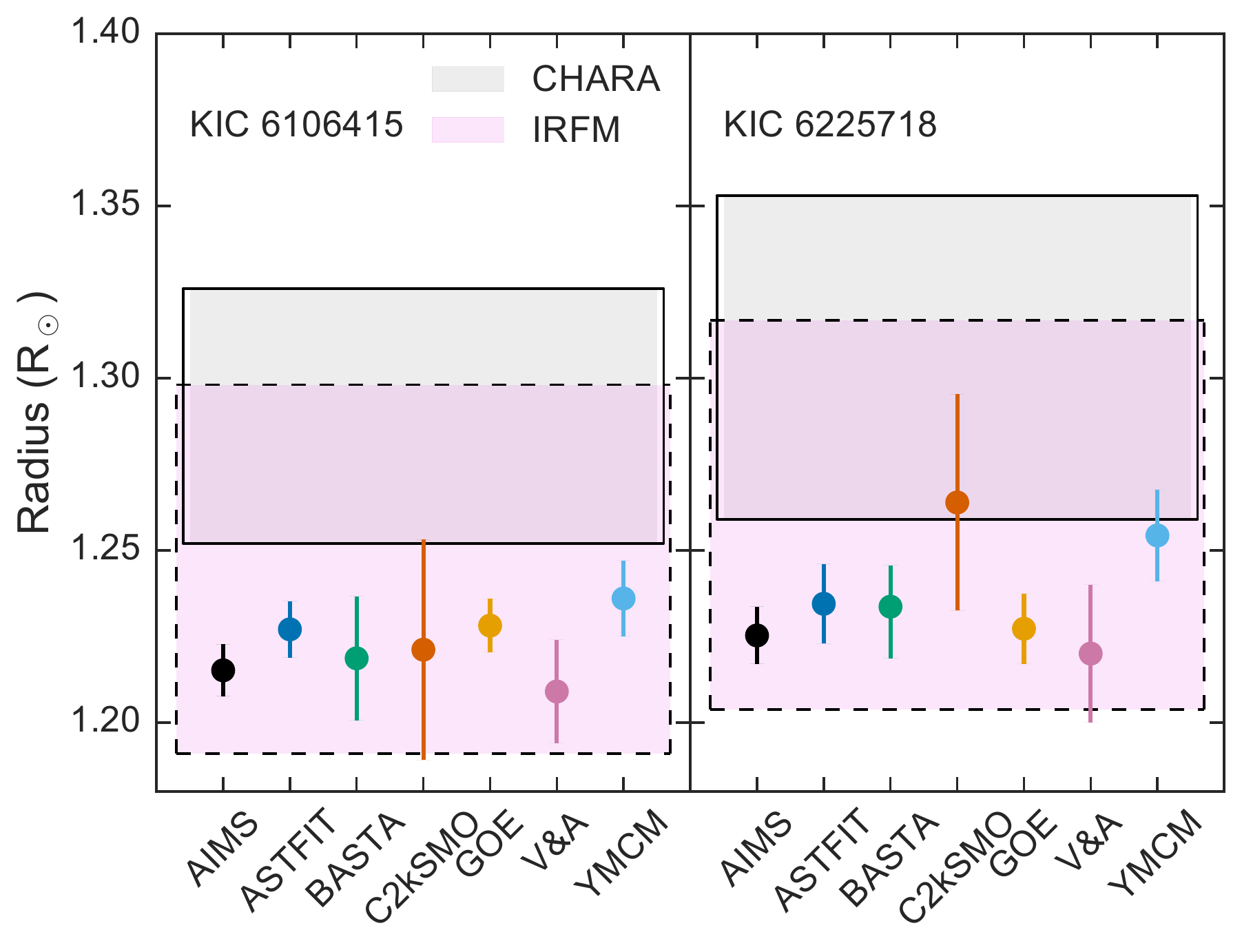}
\caption{Same as Fig.~\ref{fig:int} showing the interferometric radius from \citet{Huber:2012iv} (grey shaded region and solid lines) and the radius determined using the IRFM formulation from \citet{Casagrande:2014gx} (violet shaded region and dashed lines). See text for details.\label{fig:irfang}}
\end{figure}

Figure~\ref{fig:int} shows the results for the binary 16~Cyg A\&B \citep{White:2013bu} and KIC~8006161 \citep{Huber:2012iv}. For all three there is very good agreement between interferometric and asteroseismic radii, except for the {\ttfamily AIMS} results, which are more than 1$\sigma$ away from the observations in two cases. For the other two stars in our sample that have interferometric radii, the comparison is shown in Fig.~\ref{fig:irfang}. There is a clear difference between the radius from CHARA and the asteroseismically determined ones, the latter being systematically smaller. However, these stars have the smallest angular diameters measured in the \citet{Huber:2012iv} sample (0.289 and 0.231~mas), making them more prone to systematic errors in the adopted calibrator diameters than for larger target stars which are better resolved. In light of this, we restrict the comparison with interferometry to stars with angular diameters larger than $>0.3$~mas and also consider another independent radius determination coming from the InfraRed Flux Method (IRFM).

In the implementation considered here, the IRFM used several photometric bands to determine the bolometric flux of a star and its effective temperature in a self-consistent manner \citep[see][and Section~\ref{ss:par} below for details]{Casagrande:2010hj,Casagrande:2014gx}. A natural by-product of this procedure is the stellar angular diameter which, combined with parallax measurements, yields stellar radius as depicted in Fig.~\ref{fig:irfang}.

The agreement with asteroseismic radii is better than the comparison with interferometry, suggesting that the differences with the interferometric results indeed come from calibration problems. Additionally, the interferometric effective temperatures for these two targets are lower (by 1$-$2$\sigma$) than the IRFM and spectroscopic temperatures presented in this work and from \citet{2012MNRAS.423..122B}, translating into systematic differences of the order of $\sim$3\%. Only additional data from CHARA or another interferometer in the future will yield further information about these discrepancies, but we can conclude that the current available observations support the accuracy of asteroseismic radius determinations for these five targets. We note in passing that the IRFM results for KIC~8006161and 16~Cyg A\&B also agree well with the predictions from all pipelines.
\subsection{Parallaxes}\label{ss:par}
Using parallaxes to test the accuracy of asteroseismic results is a less direct method of comparison than e.g., radius determination from interferometry. The measured parallaxes can accurately determine the distance to our targets, which in turn can be predicted from asteroseismic radii if an estimate of the stellar angular diameter is available. Now that the Gaia mission has provided parallaxes for all stars in common with the Hipparcos and Tycho-2 catalogues \citep{2016arXiv160904303L}, we can compare our distance determinations for 64 of the targets appearing in the Gaia data release 1 \citep[DR1,][]{2016arXiv160904172G}.

We base our asteroseismic distance determinations on a method that couples radii to IRFM results in a self-consistent manner, as described by \citet{SilvaAguirre:2011es,SilvaAguirre:2012du}. Briefly, the bolometric flux (and therefore angular diameter) was estimated from Tycho and 2MASS photometry using the spectroscopic metallicity of each star and a reddening value given by the distance dependent map of \citet{2005AJ....130..659A}. We then interpolated using the asteroseismic gravity determined by each pipeline to extract a value of angular diameter and estimate its distance using the radius. We have assumed a uniform uncertainty in angular diameter of 3\% \citep[see][for further details on the implementation adopted]{Casagrande:2014bd}. The resulting asteroseismic distances have median uncertainties of the order of 4.5\%, while the median astrometric parallax uncertainty from the Gaia DR1 is close to 1\%.

\begin{figure}[ht!]
\centering
\includegraphics[width=\linewidth]{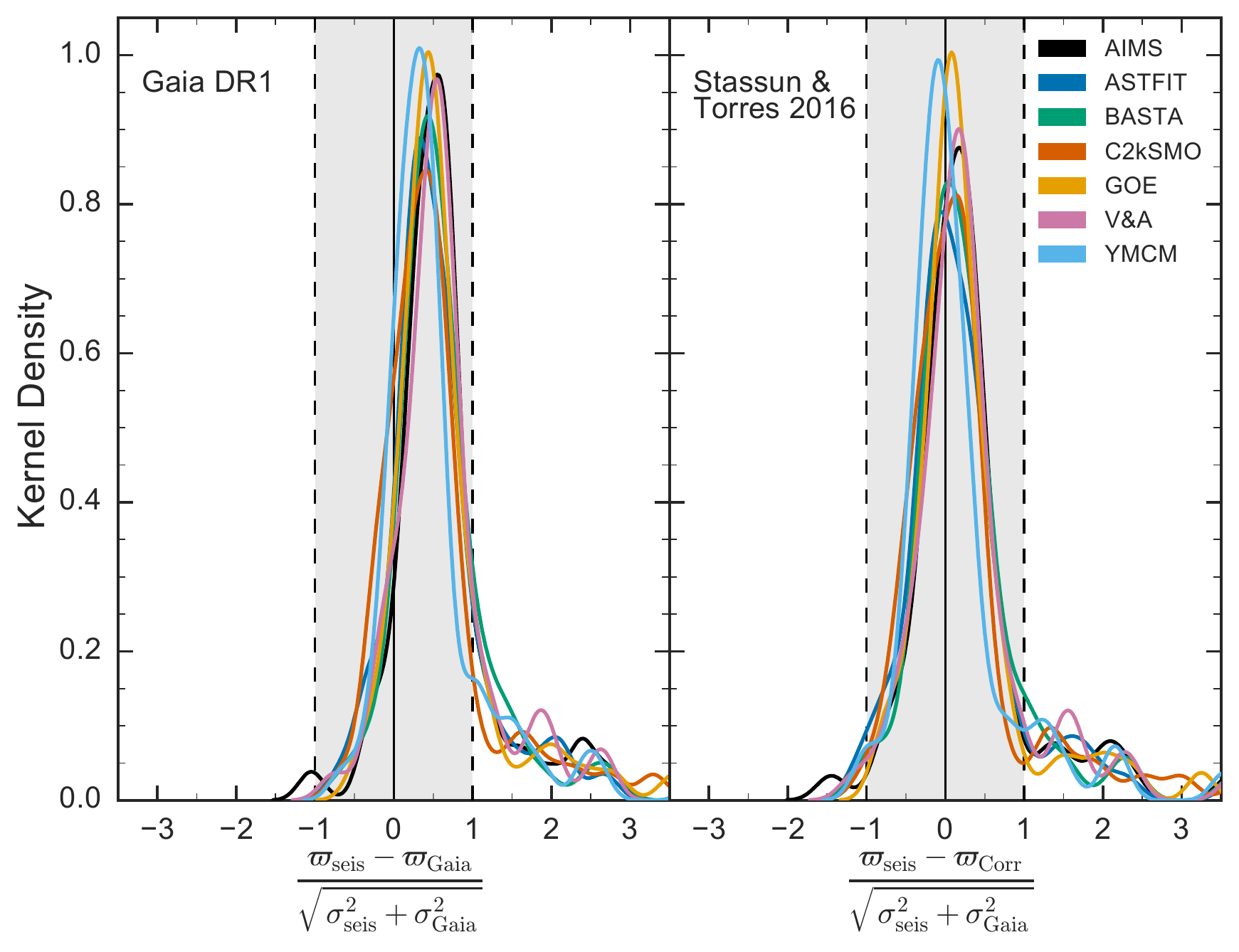}
\caption{Comparison of parallaxes determined from asteroseismic distances and those measured by the Gaia satellite. Each panel shows the distribution of the difference between both determinations normalised by the individual uncertainties. Grey shaded area and vertical dashed lines mark the 1$\sigma$ region. {\it Left:} using parallaxes directly from Gaia DR1. {\it Right:} applying the \citet{Stassun:2016ij} correction to the Gaia parallaxes. See text for details.\label{fig:par}}
\end{figure}

The resulting comparison is presented in Fig.~\ref{fig:par} where we plot the distribution of normalised differences between the asteroseismically determined distance transformed to parallax and the measurements from Gaia. There is an overall good level of agreement between all seismic pipelines, but an offset at the level of 1$\sigma$ appears when comparing to the results of DR1 (with the astrometric parallax being systematically smaller than those predicted by asteroseismology). A similar result has recently been found by \citet{Stassun:2016ij} who compared Gaia distances with independent measurements from eclipsing binaries, and found a constant offset of $-$0.25~mas in the sense of Gaia parallaxes being too small. The right panel in Fig.~\ref{fig:par} shows the comparison of our distances with those corrected by the constant offset found by \citet{Stassun:2016ij} where it is clear that the results agree much better if this systematic shift is applied. We note that 20 of our stars are also present in the Hipparcos catalogue and had asteroseismic distances available \citep[from][]{SilvaAguirre:2012du} which were recently compared to Gaia results by \citet{DeRidder:2016bl}. The authors conclude that there is excellent agreement between the new astrometric and asteroseismic parallaxes, but there is a small offset pointing towards Gaia parallaxes being too small although within the uncertainties in the determinations \citep[similar to what is found here, see Fig.~2 in][]{DeRidder:2016bl}.

Although it is tempting to corroborate the conclusion of Gaia predicting too long distances drawn by \citet{Stassun:2016ij} from our sample, we note that there are other factors that could lie at the foundation of the systematic offset observed in the distances. For instance, our results and the properties of the eclipsing binaries used by \citet{Stassun:2016ij} depend on the adopted temperature scale. If our implementation of the IRFM were to return hotter temperatures and therefore smaller angular diameters, that would result in smaller radii and therefore larger distances and could potentially solve this discrepancy. Further analysis goes beyond the scope of this paper and will be presented for larger samples in upcoming studies (Huber et al., in preparation; Sahlholdt et al., in preparation). We also mention that an uncertainty of 0.3~mas is quoted in the Gaia DR1 as systematic which is of the same order as the one found here \citep{2016arXiv160904172G}.

Despite the overall good agreement between asteroseismic and astrometric parallaxes, some outliers remain beyond the 3$\sigma$ level. Reasons that could explain this discrepancy are contaminated photometry compromising the IRFM determinations of angular diameters, inaccurate reddening estimates in the 3D dust maps, or problems with the astrometric solution used to derive parallaxes. The cases of KIC~8379927 and KIC~7510397 are examples of the latter as they are known binary stars with their fluxes affected by the presence of a companion. Similarly, the targets KIC~10454113 and KIC~9025370 have only recently been identified as spectroscopic binaries based on HARPS-N data (P.~E. Nissen, private communication). There are two other binary systems in our parallax sample, namely 16 Cyg A\&B (KIC~12069424 and KIC~12069449) and the common proper motion pair KIC~9139151 and KIC~9139163 \citep{1986A&AS...66..131H}. However, the components of these systems have wide orbital separations and can be individually resolved in photometry, so there is no reason to doubt the reliability of their parallaxes and IRFM estimates of angular diameters.

There seems to be photometric contamination in KIC~1435467 (a rotationally variable star, see \citet{Ceillier:2015gf}) and KIC~7940546 (a magnetically active star presenting flares in its light curve, see \citet{Balona:2015ka}). In the same vein, KIC~12317678 is an interesting case as it shows an infrared excess in its photometry attributed to circumstellar matter \citep{McDonald:2012cg} that could affect the fluxes used by the IRFM to extract the angular diameter. On the other hand, the parallax listed in Gaia DR1 is not consistent within the uncertainties with the value measured by the Hipparcos satellite. We expect that future data releases from Gaia will help clarify this situation. KIC~4914923 and KIC~9965715 also present deviations beyond the 3$\sigma$ level that are currently attributed to problems in the reddening estimation. The remaining stars, which account for 55 out of the 64 targets with Gaia parallaxes, have no warning flag in their photometry and show an excellent agreement in distances, confirming the reliability of asteroseismically determined radii.
\subsection{Binaries}\label{ss:bin}
One final consistency check on the seismic results can be extracted from information on known binaries in our sample. Two stars in our sample have a binary companion that has either no oscillations detected (KIC~8379927) or only global seismic parameters have been extracted in one of the components \citep[KIC~7510397,][]{Appourchaux:2015gw}. We are therefore left with two systems: KIC~12069424$-$12069449 (16 Cyg A\&B) and KIC~9139151$-$9139163. Unfortunately there are no dynamical masses derived for either of these pairs, but the fact each system was presumably formed at the same time allows us to put strong constraint on the age of the systems.

\begin{figure}[ht!]
\centering
\includegraphics[width=\linewidth]{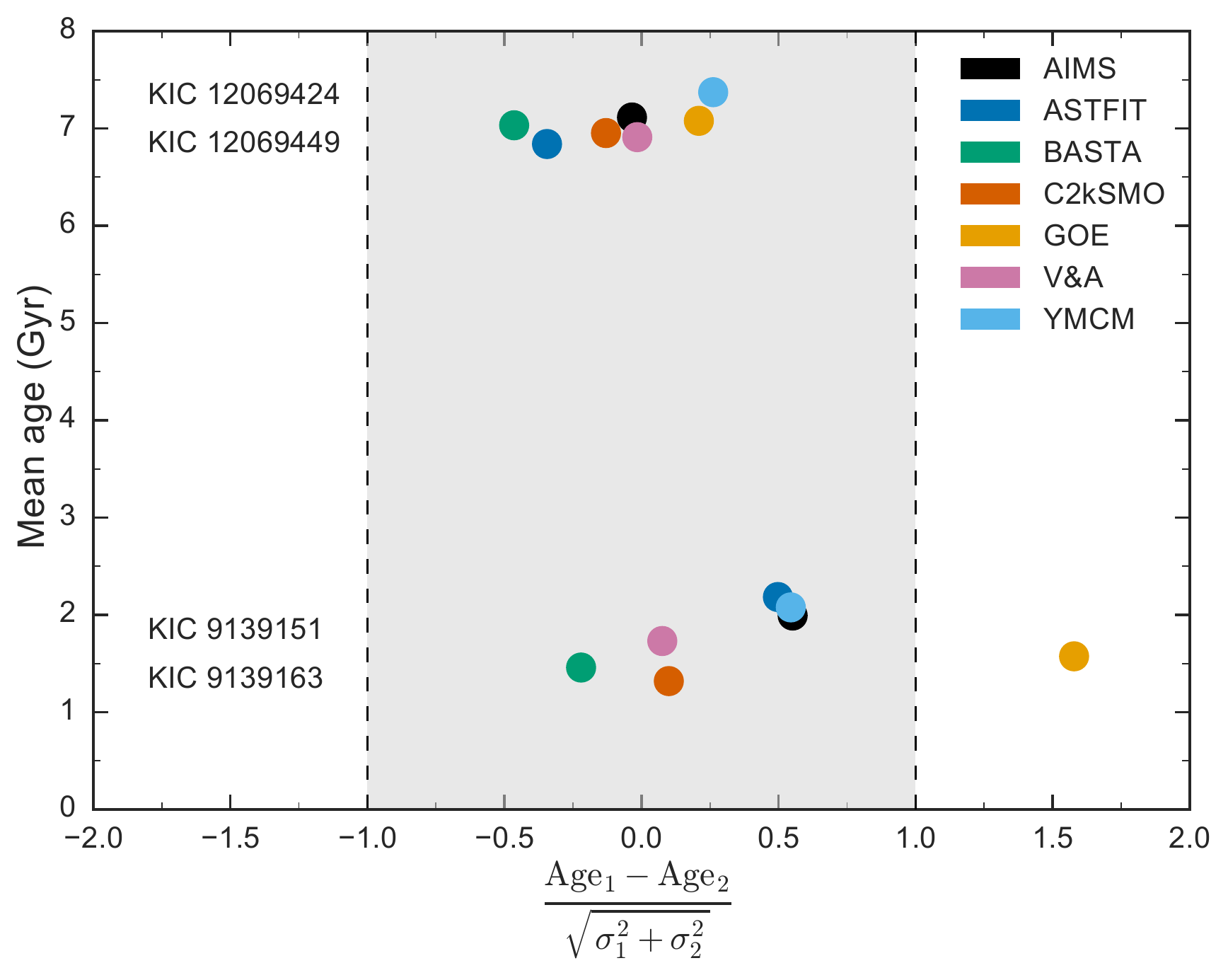}
\caption{Ages from binary stars with both components included in our sample. Plotted in the x-axis are the age differences for both components normalised by the uncertainties, while the average age of the components is plotted in the y-axis. Grey shaded region and vertical dashed lines mark the 1$\sigma$ level.\label{fig:binage}}
\end{figure}

None of the pipelines participating in this work forced the ages of these binaries to match, and Fig.~\ref{fig:binage} shows the results for the two systems included in our sample. All methods return compatible results for 16 Cyg A\&B, while {\ttfamily GOE} is the only outlier in the results for the other two stars (at the 1.5$\sigma$ level). All considered, age determinations from all pipelines are consistent for the binaries in our sample and provide further support for the accuracy of asteroseismically determined ages from individual frequencies in main-sequence stars.
\section{Conclusions and outlook}\label{s:conc}
We have presented an asteroseismic analysis for the {\it Kepler} dwarfs LEGACY sample, comprising main-sequence stars observed for more than a year for which individual oscillation frequencies have been detected. Using seven independent modelling pipelines we have determined precise physical properties for the 66 stars in the set, with average uncertainties of the order of $\sim$2\% in radius, $\sim$4\% in mass, and $\sim$10\% in age. Comparison of the results across pipelines shows a good level of agreement in the derived properties, with differences at the level expected from the variety of codes and input physics that were used in the present analysis.

Our results were subject to tests using available independent measurements of stellar properties such as those of the Sun, radii from interferometry and the IRFM, and distances from parallaxes. We found that the asteroseismically-inferred properties can reproduce the large majority of independent constraints within the uncertainties, while the deviating results can be understood in terms of stellar evolution relations (such as the mass-initial helium abundance anti-correlation) and different issues with the complementary data needed to estimate the desired property (such as contaminated photometric fluxes). These comparisons serve as confirmation of the accuracy of asteroseismic properties determined by fitting individual oscillations frequencies (or combinations thereof).

The stellar and oscillation mode properties presented here and in the accompanying paper by Lund et al.~(2016, submitted) constitute the largest sample of asteroseismically-analysed main-sequence stars using individual oscillation frequencies. Adding the stars studied by \citet{2016MNRAS.456.2183D} and \citet{2015MNRAS.452.2127S} gives 99~stars that comprise the best-characterised set of main-sequence solar-like oscillators observed by the {\it Kepler} mission. We have compared of our determined initial abundances with expectations from galactic chemical enrichment, and of our obtained efficiencies of convection with predictions from 3D hydrodynamical simulations. In both cases we found interesting features that motivate further study, such as sub-SBBN initial helium abundances and the non-dependance of the convective efficiency on stellar effective temperature. We make these results available for the community and hope they will provide the basis of extensive work going beyond our analysis, such as studies of convective envelopes \citep[][]{Mazumdar:2014kt}, surface helium abundance \citep[][]{Verma:2014fx}, core convection \citep[e.g.,][]{Deheuvels:2016ek}, inversion of the stellar interior \citep[e.g.,][]{Buldgen:2015bd,Buldgen:2015bi}, amplitudes and line widths (e.g., Houdek et al.~2016, in preparation), average 3D simulations \citep[e.g.,][]{Sonoi:2015gy, Ball:2016jl}, and many others.
\acknowledgments
Funding for this Discovery mission is provided by NASA's Science Mission Directorate. The authors acknowledge the dedicated team behind the {\it Kepler} and K2 missions, without whom this work would not have been possible. We thank Jens Jessen-Hansen for inspiration to the paper title. Funding for the Stellar Astrophysics Centre is provided by The Danish National Research Foundation (Grant DNRF106). The research was supported by the ASTERISK project (ASTERoseismic Investigations with SONG and Kepler) funded by the European Research Council (Grant agreement no.: 267864). VSA and TRW acknowledge support from VILLUM FONDEN (research grant 10118). MNL acknowledges the support of The Danish Council for Independent Research | Natural Science (Grant DFF-4181-00415). WHB acknowledges research funding by Deutsche Forschungsgemeinschaft (DFG) under grant SFB 963/1 ``Astrophysical flow instabilities and turbulence'' (Project A18). SB is partially supported by NSF grant AST-1514676 and NASA grant NNX16A109G. DRR was funded by the European Community's Seventh Framework Programme (FP7/2007-2013) under grant agreement No. 312844 (SPACEINN), thereby allowing him to develop the AIMS pipeline. WJC, and GRD acknowledge the support of the UK Science and Technology Facilities Council (STFC). D.H. acknowledges support by the Australian Research Council's Discovery Projects funding scheme (project number DE140101364) and support by the National Aeronautics and Space Administration under Grant NNX14AB92G issued through the Kepler Participating Scientist Program. This work has made use of data from the European Space Agency (ESA) mission {\it Gaia} (\url{http://www.cosmos.esa.int/gaia}), processed by the {\it Gaia} Data Processing and Analysis Consortium (DPAC, \url{http://www.cosmos.esa.int/web/gaia/dpac/consortium}). Funding for the DPAC has been provided by national institutions, in particular the institutions participating in the {\it Gaia} Multilateral Agreement.
\bibliographystyle{aasjournal}
\bibliography{legacy}
\newpage
\appendix
\section{The AIMS pipeline}\label{app:pip}
The ``Asteroseimic Inference on a Massive Scale'' (AIMS) pipeline was recently developed at the University of Birmingham in the context of the SpaceInn network. This pipeline relies on various basic components:
\begin{itemize}
\item a precalculated grid of models. In the present paper, the grid of main-sequence models from \citet{Coelho:2015ci} was used. This grid of models was calculated via the MESA \citep{Paxton:2013km,Paxton:2015iy} stellar evolution code and ranges from $0.8\,\mathrm{M}_{\odot}$ to $1.5\,\mathrm{M}_{\odot}$ in steps of $0.01\,\mathrm{M}_{\odot}$, and from $-0.8$ dex to $0.8$ dex in steps of $0.1$ dex for [Fe/H]. The \citet{Grevesse:1993vd} value of $(Z/X)_{\odot}=0.0245$ was used to convert [Fe/H] to $Z/X$. The enrichment law used $\Delta Y/\Delta Z=2$ \citep[e.g.][]{1982A&A...105..140C} and a primordial helium abundance $Y_{\mathrm{p}} = 0.2484$  \citep{Cyburt:2003ew}. Convection was based on the standard mixing length theory, using a solar-calibrated parameter. No diffusion or rotational mixing was included, while overshoot was taking into account as an adiabatic extension of the mixed region with an efficiency of  $d_\mathrm{ov}=0.20\times H_\mathrm{p}$. The NACRE thermonuclear reactions are used with the updated rate in $^{14}N(p,\gamma)^{15}O$ from \citet{Imbriani:2005he}
\item the MCMC algorithm from \citet{2013PASP..125..306F}. This provides a way of approximating the probability distribution function in stellar parameter space that results from the provided seismic and classic constraints.
\item linear interpolation within the grid of models using a Delaunay tessellation in stellar parameter space (except for the time dimension which is dealt with separately). Hence, AIMS can explore points in parameter space which lie between the stellar models from the grid. The Delaunay tessellation is calculated thanks to the qhull\footnote{See \href{http://www.qhull.org}{\tt http://www.qhull.org}} library (via python's numpy library). The advantage of using a Delaunay tessellation is that the grid of models does not necessarily have to be structured.
\end{itemize}
The AIMS pipeline is freely available\footnote{See \href{http://bison.ph.bham.ac.uk/spaceinn/aims}{\tt http://bison.ph.bham.ac.uk/spaceinn/aims}} along with a more detailed description than what is provided here.
\section{Modelling input parameters}\label{app:input}
In order to determine the stellar properties of our targets each modelling team used the individual frequencies and combinations determined by Lund et al.~2016 (submitted) in the manner described in Section~\ref{s:methods}. The necessary complementary atmospheric parameters $\teff$ and $\feh$ have been compiled from different sources in the literature and listed in Table~\ref{tab:input}, together with the global asteroseismic quantities.

Simple estimates of $\langle\dnu\rangle$ were obtained using the method of \citet{White:2011fw}, that is, from a weighted fit of the asymptotic function ($\nu_{0}(n) \simeq (n+1/2 + \epsilon)\Delta\nu$) to the radial mode frequencies from the peak-bagging as a function of radial order $n$. The weights were given by a Gaussian with a FWHM of $0.25\num$. Estimates from a more elaborate method including deviations from the simple asymptotic relation used here can be found in Lund et al.~(2016, submitted). The estimate of $\num$ was obtained from a Gaussian fit to the radial mode amplitudes from the peak-bagging.
\newpage
\begin{deluxetable*}{cllc@{$\pm$}cc@{$\pm$}cc}
\centering
\tablecaption{Global asteroseismic and atmospheric properties of our sample.\label{tab:input}}
\tablehead{
\colhead{KIC}& \colhead{$\num$ ($\mu$Hz)} &  \colhead{$\langle\dnu\rangle$ ($\mu$Hz)} &  \multicolumn2c{$\teff$ (K)}  &  \multicolumn2c{$\feh$ (dex)} &  \colhead{Reference}  
}
\startdata
1435467    &$1406.7 ^{+ 6.3   }_{ -8.4  }$&$ 70.369   ^{+ 0.034   }_{ -0.033  }$& 6326 & 77  & 0.01  & 0.1  & 1  \\ 
2837475    &$1557.6 ^{+ 8.2   }_{ -9.2  }$&$ 75.729   ^{+ 0.041   }_{ -0.042  }$& 6614 & 77  & 0.01  & 0.1  & 1  \\ 
3427720    &$2737.0 ^{+ 10.7  }_{ -17.7 }$&$ 120.068  ^{+ 0.031   }_{ -0.032  }$& 6045 & 77  & -0.06 & 0.1  & 1  \\ 
3456181    &$970.0  ^{+ 8.3   }_{ -5.9  }$&$ 52.264   ^{+ 0.041   }_{ -0.039  }$& 6384 & 77  & -0.15 & 0.1  & 1  \\ 
3632418    &$1166.8 ^{+ 3.0   }_{ -3.8  }$&$ 60.704   ^{+ 0.019   }_{ -0.018  }$& 6193 & 77  & -0.12 & 0.1  & 1  \\ 
3656476    &$1925.0 ^{+ 7.0   }_{ -6.3  }$&$ 93.194   ^{+ 0.018   }_{ -0.020  }$& 5668 & 77  & 0.25  & 0.1  & 1  \\ 
3735871    &$2862.6 ^{+ 16.6  }_{ -26.5 }$&$ 123.049  ^{+ 0.047   }_{ -0.046  }$& 6107 & 77  & -0.04 & 0.1  & 1  \\ 
4914923    &$1817.0 ^{+ 6.3   }_{ -5.2  }$&$ 88.531   ^{+ 0.019   }_{ -0.019  }$& 5805 & 77  & 0.08  & 0.1  & 1  \\ 
5184732    &$2089.3 ^{+ 4.4   }_{ -4.1  }$&$ 95.545   ^{+ 0.024   }_{ -0.023  }$& 5846 & 77  & 0.36  & 0.1  & 1  \\ 
5773345    &$1101.2 ^{+ 5.7   }_{ -6.6  }$&$ 57.303   ^{+ 0.030   }_{ -0.027  }$& 6130 & 84  & 0.21  & 0.09 & 6  \\ 
5950854    &$1926.7 ^{+ 21.9  }_{ -20.4 }$&$ 96.629   ^{+ 0.102   }_{ -0.107  }$& 5853 & 77  & -0.23 & 0.1  & 1  \\ 
6106415    &$2248.6 ^{+ 4.6   }_{ -3.9  }$&$ 104.074  ^{+ 0.023   }_{ -0.026  }$& 6037 & 77  & -0.04 & 0.1  & 1  \\ 
6116048    &$2126.9 ^{+ 5.5   }_{ -5.0  }$&$ 100.754  ^{+ 0.017   }_{ -0.017  }$& 6033 & 77  & -0.23 & 0.1  & 1  \\ 
6225718    &$2364.2 ^{+ 4.9   }_{ -4.6  }$&$ 105.695  ^{+ 0.018   }_{ -0.017  }$& 6313 & 76  & -0.07 & 0.1  & 1  \\ 
6508366    &$958.3  ^{+ 4.6   }_{ -3.6  }$&$ 51.553   ^{+ 0.046   }_{ -0.047  }$& 6331 & 77  & -0.05 & 0.1  & 1  \\ 
6603624    &$2384.0 ^{+ 5.4   }_{ -5.6  }$&$ 110.128  ^{+ 0.012   }_{ -0.012  }$& 5674 & 77  & 0.28  & 0.1  & 1  \\ 
6679371    &$941.8  ^{+ 5.1   }_{ -5.0  }$&$ 50.601   ^{+ 0.029   }_{ -0.029  }$& 6479 & 77  & 0.01  & 0.1  & 1  \\ 
6933899    &$1389.9 ^{+ 3.9   }_{ -3.6  }$&$ 72.135   ^{+ 0.018   }_{ -0.018  }$& 5832 & 77  & -0.01 & 0.1  & 1  \\ 
7103006    &$1167.9 ^{+ 7.2   }_{ -6.9  }$&$ 59.658   ^{+ 0.029   }_{ -0.030  }$& 6344 & 77  & 0.02  & 0.1  & 1  \\ 
7106245    &$2397.9 ^{+ 24.0  }_{ -28.7 }$&$ 111.376  ^{+ 0.063   }_{ -0.061  }$& 6068 & 102 & -0.99 & 0.19 & 4  \\ 
7206837    &$1652.5 ^{+ 10.6  }_{ -11.7 }$&$ 79.131   ^{+ 0.037   }_{ -0.039  }$& 6305 & 77  & 0.10  & 0.1  & 1  \\ 
7296438    &$1847.8 ^{+ 8.5   }_{ -12.6 }$&$ 88.698   ^{+ 0.040   }_{ -0.036  }$& 5775 & 77  & 0.19  & 0.1  & 1  \\ 
7510397    &$1189.1 ^{+ 3.4   }_{ -4.4  }$&$ 62.249   ^{+ 0.020   }_{ -0.020  }$& 6171 & 77  & -0.21 & 0.1  & 1  \\ 
7680114    &$1709.1 ^{+ 7.1   }_{ -6.5  }$&$ 85.145   ^{+ 0.039   }_{ -0.043  }$& 5811 & 77  & 0.05  & 0.1  & 1  \\ 
7771282    &$1465.1 ^{+ 27.0  }_{ -18.7 }$&$ 72.463   ^{+ 0.069   }_{ -0.079  }$& 6248 & 77  & -0.02 & 0.1  & 1  \\ 
7871531    &$3455.9 ^{+ 19.3  }_{ -26.5 }$&$ 151.329  ^{+ 0.025   }_{ -0.023  }$& 5501 & 77  & -0.26 & 0.1  & 1  \\ 
7940546    &$1116.6 ^{+ 3.3   }_{ -3.6  }$&$ 58.762   ^{+ 0.029   }_{ -0.029  }$& 6235 & 77  & -0.20 & 0.1  & 1  \\ 
7970740    &$4197.4 ^{+ 21.2  }_{ -18.4 }$&$ 173.541  ^{+ 0.060   }_{ -0.068  }$& 5309 & 77  & -0.54 & 0.1  & 1  \\ 
8006161    &$3574.7 ^{+ 11.4  }_{ -10.5 }$&$ 149.427  ^{+ 0.015   }_{ -0.014  }$& 5488 & 77  & 0.34  & 0.1  & 1  \\ 
8150065    &$1876.9 ^{+ 38.1  }_{ -32.4 }$&$ 89.264   ^{+ 0.134   }_{ -0.121  }$& 6173 & 101 & -0.13 & 0.15 & 4  \\ 
8179536    &$2074.9 ^{+ 13.8  }_{ -12.0 }$&$ 95.090   ^{+ 0.058   }_{ -0.054  }$& 6343 & 77  & -0.03 & 0.1  & 1  \\ 
8228742    &$1190.5 ^{+ 3.4   }_{ -3.7  }$&$ 62.071   ^{+ 0.022   }_{ -0.021  }$& 6122 & 77  & -0.08 & 0.1  & 1  \\ 
8379927    &$2795.3 ^{+ 6.0   }_{ -5.7  }$&$ 120.288  ^{+ 0.017   }_{ -0.018  }$& 6067 & 120 & -0.10 & 0.15 & 2  \\ 
8394589    &$2396.7 ^{+ 10.5  }_{ -9.4  }$&$ 109.488  ^{+ 0.034   }_{ -0.035  }$& 6143 & 77  & -0.29 & 0.1  & 1  \\ 
8424992    &$2533.7 ^{+ 27.0  }_{ -28.1 }$&$ 120.584  ^{+ 0.062   }_{ -0.064  }$& 5719 & 77  & -0.12 & 0.1  & 1  \\ 
8694723    &$1470.5 ^{+ 3.7   }_{ -4.1  }$&$ 75.112   ^{+ 0.019   }_{ -0.021  }$& 6246 & 77  & -0.42 & 0.1  & 1  \\ 
8760414    &$2455.3 ^{+ 9.1   }_{ -8.3  }$&$ 117.230  ^{+ 0.022   }_{ -0.018  }$& 5873 & 77  & -0.92 & 0.1  & 1  \\ 
8938364    &$1675.1 ^{+ 5.2   }_{ -5.8  }$&$ 85.684   ^{+ 0.018   }_{ -0.020  }$& 5677 & 77  & -0.13 & 0.1  & 1  \\ 
9025370    &$2988.6 ^{+ 20.0  }_{ -16.9 }$&$ 132.628  ^{+ 0.030   }_{ -0.024  }$& 5270 & 180 & -0.12 & 0.18 & 3  \\ 
9098294    &$2314.7 ^{+ 9.2   }_{ -10.4 }$&$ 108.894  ^{+ 0.023   }_{ -0.022  }$& 5852 & 77  & -0.18 & 0.1  & 1  \\ 
9139151    &$2690.4 ^{+ 14.5  }_{ -9.0  }$&$ 117.294  ^{+ 0.031   }_{ -0.032  }$& 6302 & 77  & 0.10  & 0.1  & 1  \\ 
9139163    &$1729.8 ^{+ 6.2   }_{ -5.9  }$&$ 81.170   ^{+ 0.042   }_{ -0.036  }$& 6400 & 84  & 0.15  & 0.09 & 6  \\ 
9206432    &$1866.4 ^{+ 10.3  }_{ -14.9 }$&$ 84.926   ^{+ 0.046   }_{ -0.051  }$& 6538 & 77  & 0.16  & 0.1  & 1  \\ 
9353712    &$934.3  ^{+ 11.1  }_{ -8.3  }$&$ 51.467   ^{+ 0.091   }_{ -0.104  }$& 6278 & 77  & -0.05 & 0.1  & 1  \\ 
9410862    &$2278.8 ^{+ 31.2  }_{ -16.6 }$&$ 107.390  ^{+ 0.050   }_{ -0.053  }$& 6047 & 77  & -0.31 & 0.1  & 1  \\ 
9414417    &$1155.3 ^{+ 6.1   }_{ -4.6  }$&$ 60.115   ^{+ 0.024   }_{ -0.024  }$& 6253 & 75  & -0.13 & 0.1  & 7  \\ 
9812850    &$1255.2 ^{+ 9.1   }_{ -7.0  }$&$ 64.746   ^{+ 0.067   }_{ -0.068  }$& 6321 & 77  & -0.07 & 0.1  & 1  \\ 
9955598    &$3616.8 ^{+ 21.2  }_{ -29.6 }$&$ 153.283  ^{+ 0.029   }_{ -0.032  }$& 5457 & 77  & 0.05  & 0.1  & 1  \\ 
9965715    &$2079.3 ^{+ 9.2   }_{ -10.4 }$&$ 97.236   ^{+ 0.041   }_{ -0.042  }$& 5860 & 180 & -0.44 & 0.18 & 3  \\ 
10068307   &$995.1  ^{+ 2.8   }_{ -2.7  }$&$ 53.945   ^{+ 0.019   }_{ -0.020  }$& 6132 & 77  & -0.23 & 0.1  & 1  \\ 
10079226   &$2653.0 ^{+ 47.7  }_{ -44.3 }$&$ 116.345  ^{+ 0.059   }_{ -0.052  }$& 5949 & 77  & 0.11  & 0.1  & 1  \\ 
10162436   &$1052.0 ^{+ 4.0   }_{ -4.2  }$&$ 55.725   ^{+ 0.035   }_{ -0.039  }$& 6146 & 77  & -0.16 & 0.1  & 1  \\ 
10454113   &$2357.2 ^{+ 8.2   }_{ -9.1  }$&$ 105.063  ^{+ 0.031   }_{ -0.033  }$& 6177 & 77  & -0.07 & 0.1  & 1  \\ 
10516096   &$1689.8 ^{+ 4.6   }_{ -5.8  }$&$ 84.424   ^{+ 0.022   }_{ -0.025  }$& 5964 & 77  & -0.11 & 0.1  & 1  \\ 
10644253   &$2899.7 ^{+ 21.3  }_{ -22.8 }$&$ 123.080  ^{+ 0.056   }_{ -0.055  }$& 6045 & 77  & 0.06  & 0.1  & 1  \\ 
10730618   &$1282.1 ^{+ 14.6  }_{ -12.7 }$&$ 66.333   ^{+ 0.061   }_{ -0.064  }$& 6150 & 180 & -0.11 & 0.18 & 3  \\ 
10963065   &$2203.7 ^{+ 6.7   }_{ -6.3  }$&$ 103.179  ^{+ 0.027   }_{ -0.027  }$& 6140 & 77  & -0.19 & 0.1  & 1  \\ 
11081729   &$1968.3 ^{+ 11.0  }_{ -12.6 }$&$ 90.116   ^{+ 0.048   }_{ -0.047  }$& 6548 & 82  & 0.11  & 0.1  & 1  \\ 
11253226   &$1590.6 ^{+ 10.6  }_{ -6.8  }$&$ 76.858   ^{+ 0.026   }_{ -0.030  }$& 6642 & 77  & -0.08 & 0.1  & 1  \\ 
11772920   &$3674.7 ^{+ 55.1  }_{ -36.1 }$&$ 157.746  ^{+ 0.032   }_{ -0.033  }$& 5180 & 180 & -0.09 & 0.18 & 3  \\ 
12009504   &$1865.6 ^{+ 7.7   }_{ -6.2  }$&$ 88.217   ^{+ 0.026   }_{ -0.025  }$& 6179 & 77  & -0.08 & 0.1  & 1  \\ 
12069127   &$884.7  ^{+ 10.1  }_{ -8.0  }$&$ 48.400   ^{+ 0.048   }_{ -0.048  }$& 6276 & 77  & 0.08  & 0.1  & 1  \\ 
12069424   &$2188.5 ^{+ 4.6   }_{ -3.0  }$&$ 103.277  ^{+ 0.021   }_{ -0.020  }$& 5825 & 50  & 0.10  & 0.026 & 5  \\ 
12069449   &$2561.3 ^{+ 5.0   }_{ -5.6  }$&$ 116.929  ^{+ 0.012   }_{ -0.013  }$& 5750 & 50  & 0.05  & 0.021 & 5  \\ 
12258514   &$1512.7 ^{+ 3.3   }_{ -2.9  }$&$ 74.799   ^{+ 0.016   }_{ -0.015  }$& 5964 & 77  & -0.00 & 0.1  & 1  \\ 
12317678   &$1212.4 ^{+ 5.5   }_{ -4.9  }$&$ 63.464   ^{+ 0.025   }_{ -0.024  }$& 6580 & 77  & -0.28 & 0.1  & 1  \\ 
\enddata
\tablecomments{All values of the average large frequency separation $\langle\dnu\rangle$ and the frequency of maximum power $\num$ have been determined by Lund et al.~2016 (submitted). Reference column indicates the source of the atmospheric parameters: (1)~\citet{Buchhave:2015cg}; (2)~\citet{Pinsonneault:2012he}; (3)~\citet{Pinsonneault:2015kd}; (4)~\citet{Casagrande:2014bd}; (5)~\citet{Ramirez:2009cb}; (6)~\citet{Chaplin:2014jf}; (7)~\citet{Huber:2013jb}.}%
\end{deluxetable*}
\section{Catalogues of stellar properties}\label{app:tab}
The resulting stellar properties determined by each pipeline are published in the online version of this paper. A description of all fields available is given in Table~\ref{tab:stelprop}. For the pipelines that do not return asymmetric uncertainties in the stellar properties we report the standard deviation in the positive and negative uncertainties column.

\onecolumngrid
\begin{deluxetable*}{ll}
\tablecaption{Stellar properties for the LEGACY dwarfs sample determined by each pipeline. All tables are available in the machine readable format in the online version of this paper. Fields are given here for guidance regarding its form and content.\label{tab:stelprop}}
\tablehead{
\colhead{Field} & \colhead{Description}
}
\startdata
{\ttfamily KIC} & {\it Kepler} Input Catalogue Identifier\\
{\ttfamily Mass} & Mass in solar units \\
{\ttfamily sMassP} & Positive mass uncertainty in solar units \\
{\ttfamily sMassM} & Negative mass uncertainty in solar units \\
{\ttfamily Rad} & Radius in solar units \\
{\ttfamily sRadP} & Positive radius uncertainty in solar units \\
{\ttfamily sRadM} & Negative radius uncertainty in solar units \\
{\ttfamily Grav} & Surface gravity in dex \\
{\ttfamily sGravP} & Positive surface gravity uncertainty in dex \\
{\ttfamily sGravM} & Negative surface gravity uncertainty in dex \\
{\ttfamily Age} & Age in units of Gyr \\
{\ttfamily sAgeP} & Positive age uncertainty in units of Gyr \\
{\ttfamily sAgeM} & Negative age uncertainty in units of Gyr \\
{\ttfamily Lum} & Luminosity in solar units \\
{\ttfamily sLumP} & Positive luminosity uncertainty in solar units \\
{\ttfamily sLumM} & Negative luminosity uncertainty in solar units \\
{\ttfamily Rho} & Density in g/cm$^3$ \\
{\ttfamily sRhoP} & Positive density uncertainty in g/cm$^3$ \\
{\ttfamily sRhoM} & Negative density uncertainty in g/cm$^3$ \\
{\ttfamily Dist} & Distance in pc \\
{\ttfamily sDistP} & Positive distance uncertainty in pc \\
{\ttfamily sDistM} & Negative distance uncertainty in pc \\
{\ttfamily Xini} & Fractional initial hydrogen abundance  \\
{\ttfamily sXiniP} & Positive fractional initial hydrogen abundance uncertainty \\
{\ttfamily sXiniM} & Negative fractional initial hydrogen abundance uncertainty \\
{\ttfamily Yini} & Fractional initial helium abundance  \\
{\ttfamily sYiniP} & Positive fractional initial helium abundance uncertainty \\
{\ttfamily sYiniM} & Negative fractional initial helium abundance uncertainty \\
{\ttfamily Xsup} & Fractional surface hydrogen abundance  \\
{\ttfamily sXsupP} & Positive fractional surface hydrogen abundance uncertainty \\
{\ttfamily sXsupM} & Negative fractional surface hydrogen abundance uncertainty \\
{\ttfamily Ysup} & Fractional surface helium abundance  \\
{\ttfamily sYsupP} & Positive fractional surface helium abundance uncertainty \\
{\ttfamily sYsupM} & Negative fractional surface helium abundance uncertainty \\
{\ttfamily Xcen} & Fractional central hydrogen abundance  \\
{\ttfamily sXcenP} & Positive fractional central hydrogen abundance uncertainty \\
{\ttfamily sXcenM} & Negative fractional central hydrogen abundance uncertainty \\
{\ttfamily Ycen} & Fractional central helium abundance  \\
{\ttfamily sYcenP} & Positive fractional central helium abundance uncertainty \\
{\ttfamily sYcenM} & Negative fractional central helium abundance uncertainty \\
{\ttfamily MCcore} & Mass coordinate of the convective core edge  \\
{\ttfamily sMCcoreP} & Positive fractional uncertainty in mass coordinate of convective core edge \\
{\ttfamily sMCcoreM} & Negative fractional uncertainty in mass coordinate of convective core edge \\
{\ttfamily Rbce} & Radius coordinate of the base of the convective envelope  \\
{\ttfamily sRbceP} & Positive fractional uncertainty in radius coordinate of the base of the convective envelope \\
{\ttfamily sRbceM} & Negative fractional uncertainty in radius coordinate of the base of the convective envelope \\
{\ttfamily $\alpha_\mathrm{conv}$} & Convective efficiency \\
{\ttfamily s$\alpha_\mathrm{conv}$P} & Positive fractional uncertainty in convective efficiency \\
{\ttfamily s$\alpha_\mathrm{conv}$M} & Negative fractional uncertainty in convective efficiency \\
{\ttfamily TAMS} & Terminal age main sequence  \\
\enddata
\tablecomments{Asymmetric uncertainties are given when available, otherwise the standard deviation of the property is given in the positive and negative uncertainty fields. Density uncertainties for the {\ttfamily C2kSMO} results are not reported, while convective efficiencies and TAMS values are not available for all pipelines (see Section~\ref{ss:pres} for details).}
\end{deluxetable*}

\end{document}